\documentclass[twocolumn,prd,superscriptaddress,preprintnumbers,nofootinbib]{revtex4-2}

\usepackage{comment} 

\usepackage{graphicx}
\usepackage{amsmath,bm,amssymb,amsfonts,dsfont}
\usepackage[usenames,dvipsnames]{xcolor}
\usepackage[normalem]{ulem}
\usepackage{url}
\usepackage{array}
\usepackage{booktabs}
\usepackage{multirow}
\usepackage{float}
\usepackage[colorlinks  = true,
            linkcolor   = NavyBlue,
            urlcolor    = NavyBlue,
            citecolor   = NavyBlue,
            anchorcolor = NavyBlue]{hyperref}
\usepackage{cprotect}

\usepackage{appendix}

\usepackage[switch]{lineno}

\usepackage{tikz-feynman}
\tikzfeynmanset{compat=1.1.0}

\pdfoutput=1 

\newcommand{\lsim}{\mathrel{\mathop{\kern 0pt \rlap
  {\raise.2ex\hbox{$<$}}}
  \lower.9ex\hbox{\kern-.190em $\sim$}}}
\newcommand{\gsim}{\mathrel{\mathop{\kern 0pt \rlap
  {\raise.2ex\hbox{$>$}}}
  \lower.9ex\hbox{\kern-.190em $\sim$}}}

\tikzfeynmanset{warn luatex=false} 

\begin{document}



\title{Leptophilic dark matter in $U(1)_{L_{i}-L_{j}}$ models: a solution to the {\sc Fermi-Lat} Galactic Center Excess consistent with cosmological and laboratory observations}

\author{Jordan~Koechler}
\email{jordan.koechler@gmail.com}
\affiliation{Istituto Nazionale di Fisica Nucleare, Sezione di Torino, Via P. Giuria 1, 10125 Torino, Italy}

\author{Mattia Di Mauro}
\email{dimauro.mattia@gmail.com}
\affiliation{Istituto Nazionale di Fisica Nucleare, Sezione di Torino, Via P. Giuria 1, 10125 Torino, Italy}

\begin{abstract}
The particle origin of dark matter (DM) remains elusive despite decades of direct, indirect, and collider searches. Several groups have reported a $\gamma$-ray excess toward the Galactic Centre—commonly referred to as the Galactic Centre Excess (GCE). Its spectrum is consistent with annihilation of weakly interacting massive particles (WIMPs) of mass $\mathcal{O}(10\text{–}100)\,\mathrm{GeV}$ and a thermal-relic cross section. Although many concrete WIMP models reproduce the GCE spectrum, most are now excluded by direct detection experiments that are approaching the neutrino floor. We investigate a class of anomaly-free extensions of the Standard Model featuring gauged differences of lepton number, $U(1)_{L_i-L_j}$, and gauged baryon minus lepton number, $U(1)_{B-L}$. 
We show that these models can reproduce the GCE while remaining compatible with the observed relic abundance. We then impose collider and direct detection constraints, accounting for both tree-level and loop-induced kinetic mixing. The $L_\mu-L_e$ model gives the best fit to the GCE: a DM mass of $m_\chi\sim 40\text{–}50\,\mathrm{GeV}$ remains consistent with the muon and electron magnetic moment anomalies, $(g-2)_{\mu,e}$, as well as current collider and direct detection limits, for mediator masses in the range $m_{A'}\sim 70\text{–}86\,\mathrm{GeV}$ and a DM–mediator coupling of $(1\text{–}5)\times10^{-2}$. By contrast, the $L_e-L_\tau$ and $L_\mu-L_\tau$ models yield poorer fits; satisfying both the relic density and experimental bounds forces the DM mass to lie very close to resonance (\textit{i.e.}~approximately half the mediator mass). Finally, while the $B-L$ model also matches the GCE well, its parameter space is almost entirely ruled out by strong direct detection limits, except for the narrow resonance region where $m_\chi$ should be equal to $m_{A'}/2$ requiring a fine-tuning at the few-percent level.

\end{abstract}


\maketitle

\flushbottom

\section{Introduction}

The nature of dark matter (DM) remains one of the main puzzles in physics. Gravitational evidence observed from astrophysical objects at different scales---from galaxy clusters down to individual galaxies---can be interpreted by invoking at least one new particle beyond the Standard Model of particle physics (SM)~\cite{Cirelli:2024ssz}.

The main strategies for searching for DM interactions in laboratories are as follows. Indirect detection seeks the products of DM annihilation or decay in space. Among all possible SM final states, antimatter (positrons, antiprotons and antinuclei) and high‑energy photons ($\gamma$-rays) are typically considered because they are the rarest components of cosmic rays \cite{Gaskins:2016cha}. Direct detection aims to measure the recoil energy produced by Galactic halo DM particles scattering elastically off detector nuclei, which are often noble gases \cite{Schumann:2019eaa}. Finally, collider experiments (for example, at the Large Hadron Collider) attempt to produce DM particles through collisions of SM particles—such as $e^+e^-$ or $pp$—and infer their presence via missing transverse energy or momentum in the reconstructed events \cite{Boveia:2018yeb}.

Indirect detection is particularly interesting because it probes DM in its natural habitat—the Universe—where its presence is already established through gravitational effects. As noted above, indirect searches focus on the fluxes of cosmic positrons, antiprotons, antinuclei, and electromagnetic radiation from radio wavelengths up to $\gamma$-rays (see \textit{e.g.}, Refs.~\cite{DiMauro:2015jxa,DiMauro:2015tfa,Cuoco:2017rxb,Genolini:2021doh,Genolini:2021doh,DiMauro:2021qcf,McDaniel:2023bju,Balan:2023lwg}). In particular, $\gamma$-rays travel in straight lines, unaffected by magnetic fields, allowing telescopes to be pointed directly at regions predicted to have the highest DM density. In this regard, directions near the center of the Milky Way are expected to yield the strongest $\gamma$‑ray flux on Earth from DM annihilation or decay \cite{Pieri:2009je}.

Several independent studies have reported an excess of $\gamma$‑ray emission detected by the {\sc Fermi} Large Area Telescope ({\sc Fermi-Lat}) toward the Galactic Center (GC) (see \textit{e.g.}, Refs.~\cite{Goodenough:2009gk,Hooper:2010mq,Boyarsky:2010dr,Hooper:2011ti,Abazajian:2012pn,Gordon:2013vta,Abazajian:2014fta,Daylan:2014rsa,Calore:2014nla,Calore:2014xka,TheFermi-LAT:2015kwa,TheFermi-LAT:2017vmf,DiMauro:2019frs,DiMauro:2021raz,Cholis:2021rpp}). This feature, known as the Galactic Center Excess (GCE), has been observed under various background models—including point sources, extended sources, interstellar emission (IEM), {\sc Fermi} bubbles, and isotropic components—and with diverse data selections and analysis techniques. 

Refs.~\cite{DiMauro:2021raz,Cholis:2021rpp} (respectively referenced Di Mauro+21 and Cholis+22 hereafter) have recently provided among the most accurate characterization of the spectrum and morphology of the GCE to date by using the template fitting approach. Their analysis confirms that the spectral energy distribution (SED) peaks at a few GeV and extends up to $\sim50$ GeV, with normalization variations of up to 60\% across different IEMs and analysis choices. The spatial morphology follows a generalized Navarro–Frenk–White (NFW) profile with slope $\gamma=1.2–1.3$, which does not vary with photon energy.

The origin of the GCE remains uncertain. Refs.~\cite{Bartels:2015aea,Lee:2015fea} applied wavelet analyses and non‑Poissonian template fitting to {\sc Fermi-Lat} data and found evidence for a faint population of unresolved sources near the GC, potentially millisecond pulsars in the bulge. This interpretation is reinforced by Refs.~\cite{Macias:2016nev,Bartels:2017vsx,Manconi:2024tgh}, who used stellar bulge models—combining nuclear and boxy components—as pulsar tracers. Their model provides a better fit than those using smooth DM spatial templates, suggesting the GCE may not be perfectly spherically symmetric. However, recent studies \cite{Leane:2019uhc,Chang:2019ars} have shown that non‑Poissonian template fitting can misattribute modeling imperfections in the IEM to spurious point‑source signals. Together with the results of Refs.~\cite{Zhong:2019ycb,Calore:2021jvg,List:2025qbx}, these findings cast significant doubt on the robustness of the pulsar interpretation.

The characteristics of the GCE are consistent with a DM origin. A spherically symmetric DM halo centered on the GC naturally produces an energy‑independent morphology, and the observed SED aligns with expectations for DM annihilation into $b\bar b$ at a thermal cross section \cite{Daylan:2014rsa,Calore:2014nla,Aghanim:2018eyx}. 

If DM is responsible for the GCE, 
$\gamma$‑ray signals with a similar spectral shape should also be detectable from the dwarf spheroidal galaxies (dSphs) in the Local Group, which have high mass‑to‑light ratios ($\sim100$–$1000$) and very low astrophysical backgrounds
\cite{Abdo_2010}. 
Yet dedicated searches toward known dSphs
(\textit{e.g.}, Refs.~\cite{Abdo_2010,Ackermann:2015zua,Lopez:2015uma,Fermi-LAT:2016uux,%
Calore:2018sdx,Hoof:2018hyn,2019MNRAS.482.3480P,McDaniel:2023bju})
have found no significant $\gamma$‑ray emission, imposing strong constraints on 
the DM interpretation of the GCE. 
Nonetheless, the current uncertainties in the DM density both near the 
GC and in dSphs are such that the non‑detection in the latter does not yet rule out a DM origin for the GCE \cite{DiMauro:2021qcf}.

If DM particles annihilate with a significant branching ratio into
hadronic final states—such as quarks and SM gauge or Higgs
bosons—they can also generate a flux of cosmic antiprotons detectable by
{\sc Ams‑02}. Constraints derived from the antiproton flux are particularly
stringent for diffusive halo half‑heights $L\lesssim\text{few}\,\mathrm{kpc}$
(see \textit{e.g.}, Ref.~\cite{DiMauro:2021qcf}). Even for values of $L$ between $4$–$8\,\mathrm{kpc}$, which are
favoured by cosmic‑ray beryllium isotopes, radio and $\gamma$‑ray observations,
the DM explanation of the GCE is strongly constrained by the observed \textsc{Ams-02} $\bar{p}$ flux
\cite{DiMauro:2021qcf,Calore:2022stf}.

The GCE can be well fitted with different theories beyond the SM (BSM), from simplified models up to UV-complete theories. One of the simplest is the singlet Higgs portal model, where DM is a singlet scalar that communicates with SM particles through the SM Higgs. This model can nicely fit the GCE and remain compatible at the same time with collider, direct detection, and limits from the non-detection of dSphs, but only in a very narrow region around the resonance, \textit{i.e.}, when the DM mass is about half the SM Higgs mass \cite{DiMauro:2023tho}. 
Most weakly interacting massive particles (WIMPs) that couple at tree-level to quarks share the same fate: they are excluded by orders of magnitude in most of the parameter space by direct detection experiments.

In this paper, and for the first time, we test the compatibility of models that gauge global accidental symmetries of the SM with the GCE and other DM searches. In particular, we consider BSM theories that gauge differences of leptonic numbers. These models are anomaly-free without requiring the addition of extra particles (see \textit{e.g.}, Refs.~\cite{Foot:1990mn,1991PhRvD..43...22H,Heeck:2011wj,Bauer:2018onh}):
\begin{equation}
\label{eq:U1models}
\mathrm{U}(1)_{L_\mu - L_e},\quad \mathrm{U}(1)_{L_e - L_\tau},\quad \mathrm{U}(1)_{L_\mu - L_\tau}.
\end{equation}
The difference between baryon and lepton numbers, $\mathrm{U}(1)_{B-L}$, is also anomaly-free if right-handed neutrinos are introduced. 
In $L_i - L_j$ models, DM particles annihilate at tree-level into leptons, while annihilation into quarks occurs only via kinetic mixing, which collider experiments constrain to be very small. Therefore, these leptophilic models naturally evade the tight constraints from the \textsc{Ams-02} $\bar{p}$ flux and are far less constrained by direct detection bounds than standard WIMP models.
We also consider the model $U(1)_{B-L}$ that gauges the difference of baryonic and leptonic numbers ($B-L$). In this BSM DM particles couple to quarks at tree-level and thus most of the parameter space is ruled out by direct detection.

Among the models listed in Eq.~\ref{eq:U1models}, $\mathrm{U}(1)_{L_\mu - L_\tau}$ has been widely studied in recent years for its potential to explain hints of BSM physics at the $4\sigma$ level in the anomalous muon magnetic moment $a_{\mu}=(g-2)_{\mu}/2$, which represents one of the most intriguing discrepancies between experimental measurements and theoretical predictions in particle physics. Recent precise measurements by the {\sc Muon} $g-2$ collaboration at Fermilab \cite{Abi:2021gix,Muong-2:2023cdq} confirm earlier findings from Brookhaven National Laboratory~\cite{Bennett:2006fi}, indicating a deviation from the SM prediction at the level of approximately $5\sigma$. 
However, a recent reevaluation of the SM $a_{\mu}=(g-2)_{\mu}/2$, with improved estimates of the QED, electroweak, and hadronic contributions, leads to a value now compatible with the Fermilab measurement \cite{Aliberti:2025beg}. Therefore, no clear signal of new physics seems to be present.

In this paper, we first evaluate the values of the $L_i-L_j$ model parameters (DM mass and couplings) that fit the GCE energy spectrum. Then we compute the upper limits on the model parameters coming from direct detection via tree-level and loop-induced kinetic mixing. Finally, we include constraints from collider searches for the dark photon and the latest limits from the anomalous muon and electron magnetic moments ($a_{\mu}$ and $a_{e}$) and show what part of the parameter space can fit the GCE spectrum remaining compatible with the other DM searches at the same time.

This paper is organized as follows. In Sec.~\ref{sec:model} we present the Lagrangian, define the relevant fields and couplings, and discuss kinetic mixing. In Sec.~\ref{sec:observables} we summarize the constraints on the model from (i) direct detection—including tree-level and loop-induced kinetic mixing—(ii) collider experiments, (iii) the relic abundance, and (iv) indirect detection using the GCE spectrum, the \textsc{Ams-02} positron flux, and upper limits from \textsc{Fermi-Lat} observations of dwarf spheroidal galaxies (dSphs). In Sec.~\ref{sec:results} we present the GCE fit and the combined constraints from relic density, collider searches, and direct detection. Finally, in Sec.~\ref{sec:conclusions} we draw our conclusions.

\medskip

\section{$L_i-L_j$ and $B-L$ Models}
\label{sec:model}

\subsection{Introduction to the model}

    
    

We consider in this paper four different BSMs that gauge the accidental global symmetries of the SM.

The first three take into account the difference of the leptonic numbers $U(1)_{L_{i}-L_{j}}$\footnote{We will label the SM leptons that are charged under the group $U(1)_{L_{i}-L_{j}}$ with $i$ and $j$ indices, while the third family that is neutral is labeled with $k$. For example, for the $U(1)_{L_{\mu}-L_{\tau}}$, $i=\mu$, $j=\tau$, and $k=e$.}, with a new gauge boson $X_\mu$, sometimes labeled as the dark photon, and a fermionic Dirac DM particle $\chi$ with mass $m_{\chi}$. The Lagrangian of the model is given by:
\begin{eqnarray}
\label{eq:lagr}
\mathcal{L} &=& \mathcal{L}_\text{SM} -\frac{1}{4}X_{\mu\nu}X^{\mu\nu} - \frac{\epsilon}{2} F_{\mu\nu} X^{\mu\nu} + \frac{1}{2} m_{X}^2 X_\mu X^\mu + \nonumber \\
&-& g_X \sum_{n=i,j}q_{X,n}(\bar{\ell}_n\gamma_\mu \ell_n + \bar{\nu}_n\gamma_\mu \nu_n) X^\mu + \\
&-& m_{\chi} \bar{\chi}\chi - g_X q_{X} \bar{\chi} \gamma_{\mu} \chi X^{\mu} \nonumber,
\end{eqnarray}
where $\mathcal{L}_\text{SM}$ is the SM Lagrangian, and the sum is performed for $i \neq j$ over $n= (e, \mu)$, $(e,\tau)$ or $(\mu,\tau)$. $\ell_n$ are the charged leptons and $\nu_n$ are the corresponding neutrinos. $F_{\mu\nu}$ is the field strength tensor of the hypercharge gauge boson $B_\mu$. $X_{\mu\nu}$, $g_X$, and $m_X$ are the field strength tensor, coupling, and mass term of the new $U(1)$ gauge boson $X_{\mu}$, respectively. 
The leptons $\ell_i/\nu_i$ and $\ell_j/\nu_j$ have opposite $U(1)_{L_{i}-L_{j}}$ charges ($q_{X,i} = -q_{X,j} = 1$), while the DM particle charge is $q_X$. We will take $q_X = 1$ unless otherwise stated.
The $X$ and DM masses can arise from either a Higgs or a Stueckelberg mechanism. In the former case, additional effects due to the extra Higgs boson are present but not considered in this paper.

In general, kinetic mixing between non-abelian and abelian gauge groups is not possible at the renormalizable level \cite{Burgess:2008ri}. However, if a non-abelian gauge group is broken as $SU(N) \rightarrow U(1)$ at some high scale, loop effects from fields charged under both this new $U(1)$ and $U(1)_Y$ induce a kinetic mixing parameter in the broken phase \cite{Burgess:2008ri}. 
We include in the model the kinetic mixing between $B_{\mu}$ and $X_{\mu}$, which is parametrized through the parameter $\epsilon$.

The fields $B_\mu$ and $X_\mu$ indicated in Eq.~\ref{eq:lagr} are not canonically normalized, as the kinetic terms are non-diagonal. Furthermore, they are not mass eigenstates.
After a proper rotation of the fields (see Appendix \ref{app:massstates} for the full calculation), we obtain the mass eigenstates associated with the photon $A_{\mu}$, the neutral electroweak boson $Z_{\mu}$, and the new gauge boson $A'_{\mu}$. Assuming a small kinetic mixing $\epsilon\ll 1$, the interaction Lagrangian becomes \cite{Bauer:2018onh}:
\begin{eqnarray}
\label{eq:intL}
&&\mathcal{L}_{\text{int}} \approx - e A_\mu J_{\rm EM}^\mu - Z_\mu [g_Z \;J_Z^\mu + g_X \sin\xi  J_X^\mu] \\
&-& A'_\mu \Big[g_X\,  J_X^\mu \;-\; e \epsilon \cos\theta_W  J_{\rm EM}^\mu  + g_Z( \epsilon \sin \theta_W  -  \sin\xi )J_Z^\mu ]. \nonumber
\end{eqnarray}
Here, $\xi \simeq \frac{\epsilon\sin\theta_W}{1-\delta}$, $\delta \equiv m_X^2/m_{Z}^2$, $\theta_W$ is the Weinberg angle, $J_Y$ is the electromagnetic current, $J_Z$ is the electroweak neutral current, and $J_X$ is the current of $U(1)_{L_i-L_j}$ defined as $J^{\mu}_X = \bar{\ell}_i\gamma^\mu \ell_i + \bar{\nu}_i\gamma^\mu \nu_i - \bar{\ell}_j\gamma^\mu \ell_j - \bar{\nu}_j\gamma^\mu \nu_j + q_X \bar{\chi} \gamma^{\mu}\chi$.
For $m_X \ll m_Z$ the mixing between $A'$ and $J_Z$ becomes negligible ($\epsilon \sin \theta_W \approx \sin \xi$).

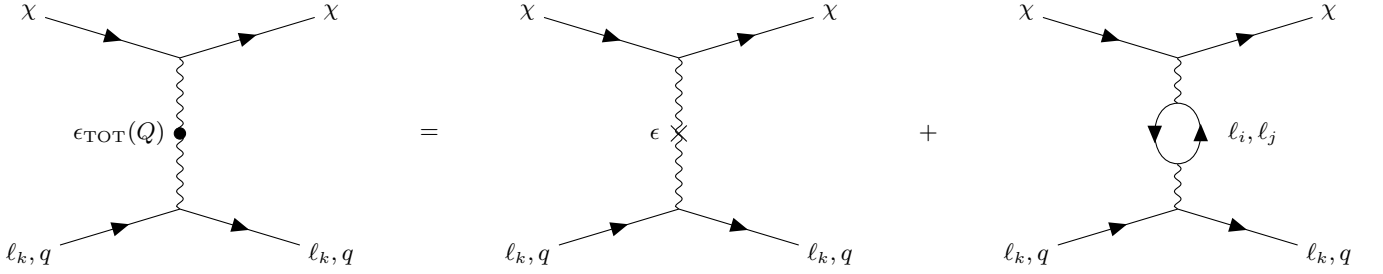
\begin{figure*}[]
  \centering
  \begin{tikzpicture}[baseline=(current bounding box.center)]
    \begin{feynman}
      \vertex (a1) at (-2,  1.6) {$\chi$};
      \vertex (a2) at (-2, -1.6) {$\ell_k,q$};
      \vertex (b1) at ( 2,  1.6) {$\chi$};
      \vertex (b2) at ( 2, -1.6) {$\ell_k,q$};
      \vertex (c1) at ( 0,  1.0);
      \vertex (c2) at ( 0, -1.0);
      \vertex[dot, label=left:$\epsilon_{\text{TOT}}(Q)$] (ins) at (0,0) {};
      \diagram*{
        (a1) -- [fermion] (c1) -- [fermion] (b1),
        (a2) -- [fermion] (c2) -- [fermion] (b2),
        (c1) -- [boson]   (c2),
      };
    \end{feynman}
    \node at (3.3,0) {$=$};
    \begin{scope}[shift={(6.6,0)}]
      \begin{feynman}
        \vertex (a1) at (-2,  1.6) {$\chi$};
        \vertex (a2) at (-2, -1.6) {$\ell_k,q$};
        \vertex (b1) at ( 2,  1.6) {$\chi$};
        \vertex (b2) at ( 2, -1.6) {$\ell_k,q$};
        \vertex (c1) at ( 0,  1.0);
        \vertex (c2) at ( 0, -1.0);
        \diagram*{
          (a1) -- [fermion] (c1) -- [fermion] (b1),
          (a2) -- [fermion] (c2) -- [fermion] (b2),
          (c1) -- [boson, insertion=0.5, edge label'=$\epsilon$, inner sep=7pt] (c2),
        };
      \end{feynman}
    \end{scope}
    \node at (9.9,0) {$+$};
    \begin{scope}[shift={(13.2,0)}]
      \begin{feynman}
        \vertex (nuL) at (-2,  1.6) {$\chi$};
        \vertex (nuR) at ( 2,  1.6) {$\chi$};
        \vertex (eL)  at (-2, -1.6) {$\ell_k,q$};
        \vertex (eR)  at ( 2, -1.6) {$\ell_k,q$};

        \vertex (t)    at (0,  1.0);
        \vertex (mid1) at (0,  0.4);
        \vertex (mid2) at (0, -0.4);
        \vertex (b)    at (0, -1.0);

        \diagram*{
          (nuL) -- [fermion] (t) -- [fermion] (nuR),
          (eL)  -- [fermion] (b) -- [fermion] (eR),

          (t)    -- [boson] (mid1),
          (mid2) -- [boson] (b),

          (mid1) -- [fermion, half right, looseness=1.3] (mid2)
                 -- [fermion, half right, looseness=1.3] (mid1),
        };
        \node[black] at (1.0,0) {$\ell_i,\ell_j$};
      \end{feynman}
    \end{scope}
  \end{tikzpicture}
  \caption{Loop–induced kinetic mixing for the scattering of the DM particle $\chi$ with a quark $q$ or the $\ell_k$ lepton family: the total form factor \(\epsilon_{\text{TOT}}(Q)\) (left panel) decomposes into the tree-level insertion \(\epsilon\) (middle panel) plus the $\ell_i$ and $\ell_j$ loop vacuum polarisation (right panel).}
  \label{fig:feynmdiag}
\end{figure*}

Looking at the interaction Lagrangian in Eq.~\ref{eq:intL}, we note a few important aspects of the theory:
\begin{itemize}
\item The vertex between the DM particle $\chi$ (or the $i,j$ SM leptons) and the $A'_\mu$ boson is proportional to $g_X$. The dominant process for achieving the correct relic density $\Omega_\chi h^2$ is DM annihilating into $i,j$ leptons (see Sec.~\ref{sec:relic}). Therefore, as dictated by the WIMP scenario, $g_X$ is expected to be of the order of the electroweak couplings $g$ and $g'$ in order to match $\Omega_\chi h^2$.
\item The new boson $A'$ can decay into quarks and into the lepton family $k$ with a vertex proportional to $\epsilon$. $A'$ dark photon decays into fermions are probed by collider experiments (see Sec.~\ref{sec:collider}), which place strong upper limits on the kinetic mixing parameter $\epsilon$, typically at the level of $\epsilon \lesssim 10^{-3}$ for $m_{A'} > 0.1$ GeV (see, \textit{e.g.}, \cite{Bauer:2018onh}). This implies that the interaction vertex between $A'$ and the lepton family $k$ or quarks is suppressed by at least 5 orders of magnitude compared to the vertex between $A'_{\mu}$ and $i,j$ families (\textit{i.e.}, $\propto \epsilon^2/g_X^2$).
\item The direct detection process proceeds via the $t$-channel diagram in Fig.~\ref{fig:feynmdiag}, which is proportional to $(g_X g \epsilon)^2$. Therefore, compared to standard WIMP models without kinetic mixing, the direct detection signal is suppressed by a factor $\epsilon^2$.
\end{itemize}
In this model we expect to achieve the correct relic density with values of $g_X$ of the order of $g/g'$, while avoiding the strong direct detection limits that affect standard WIMP models that couples at tree-level with quarks.

We also take into account a model that adds to the SM a $U(1)_{B-L}$. In this model leptons have a charge $q_X=-1$ while quarks take $q_X=+1/3$.
The Lagrangian of this model is described as in Eq.~\ref{eq:lagr} except that the the current $J_X$ is described by:
\begin{equation}
    J^{\mu}_X = \frac{1}{3}  \sum_i \bar{q}_i\gamma^\mu q_i-\sum_j \bar{\ell}_j\gamma^\mu \ell_j + q_X \bar{\chi} \gamma^{\mu}\chi,
\end{equation}
where the two sums are performed over $i$ ($j$) that represents the number of quarks (leptons). Therefore, in this model the new gauge boson $X_\mu$ couples to the quarks with a vertex proportional to $g_X$ implying that the interaction cross section of DM with quarks is proportional to $g_X^4$, \textit{i.e.}~it is not suppressed by the kinetic mixing as for $L_i-L_j$.

For each $U(1)_{L_i-L_j}$ and $U(1)_{B-L}$ model we build the associated {\tt UFO} \cite{Degrande:2011ua} and {\tt CalcHEP} files \cite{Belyaev:2012qa} by using {\tt FeynRules} \cite{Alloul:2013bka}. Then we provide the files to {\tt MadDM} \cite{Backovic:2013dpa,Ambrogi:2018jqj,Arina:2021gfn} and {\tt micrOMEGAs} \cite{Belanger:2006is,Belanger:2013oya,Belanger:2018ccd,Alguero:2023zol} to evaluate the signal for indirect detection and relic density\footnote{We use version {\tt 6.0.4} of {\tt micrOMEGAs} and {\tt 3.2} of {\tt MadDM}.}. For direct detection instead we will use our own implementation (see Sec.~\ref{sec:direct}).

\subsection{Kinetic mixing}

$U(1)_{L_i - L_j}$ models feature suppressed interactions with the generation $k$ of SM leptons and quarks that are proportional to the kinetic mixing $\epsilon$. 
Nevertheless, since the leptons $i,j$ are also charged under the electromagnetic $U(1)$, there exists an unavoidable kinetic mixing at the loop level (see right panel of Fig.~\ref{fig:feynmdiag}). This allows us to probe these BSMs also in experiments with generation $k$ of leptons and with quarks. Let us now consider this loop-induced kinetic mixing in more detail.

The kinetic mixing received contribution not only from the Lagrangian term $\epsilon F_{\mu\nu} X^{\mu\nu}/2$ but also from radiative corrections with virtual $\ell_i$ and $\ell_j$ lepton exchange at loop level.
In Fig.~\ref{fig:feynmdiag} we report the total contribution for direct detection scattering between a DM particle $\chi$ and the quark $q$.

In particular, taking into account both contributions the total kinetic mixing parameters $\epsilon_{\rm{TOT}}(Q)$ becomes~\cite{Hapitas:2021ilr}:
\begin{equation}
\label{eq:epsilontot}
\epsilon_\textrm{TOT}(Q) = \epsilon - \frac{e \, g_{X}}{2\pi^2} \int_0^1 dx\, x(1 - x) \log \left( \frac{m_j^2 + Q^2 x(1 - x)}{m_i^2 + Q^2 x(1 - x)} \right),
\end{equation}
where $Q$ is the momentum transfer of the process, \textit{i.e.}~in direct detection. 
Eq.~\ref{eq:epsilontot} should also include the contribution from left-handed neutrinos that are massive ($m_\nu$) and whose expression of loop-induced $\epsilon$ is similar with respect to the one of $\ell_{i,j}$. However, since $Q\gg m_\nu$ the contribution from neutrinos is negligible.

\begin{figure}
\centering
\includegraphics[width=0.99\linewidth]{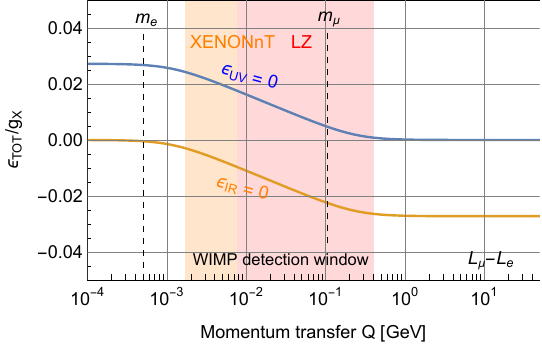}
\includegraphics[width=0.99\linewidth]{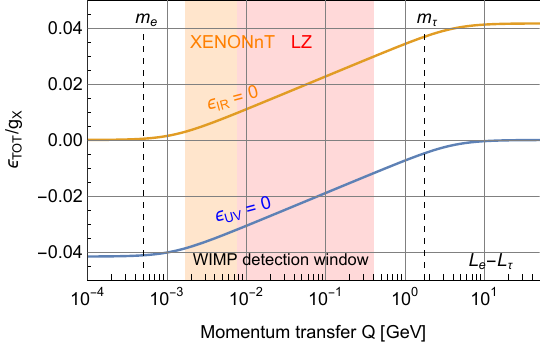}
\includegraphics[width=0.99\linewidth]{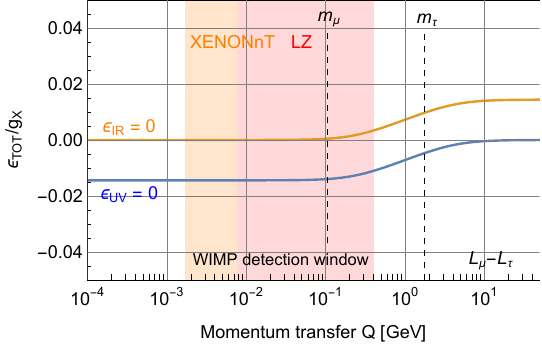}
    \caption{Momentum dependent kinetic mixing $\epsilon_{\rm{TOT}}(Q)$ divided by the coupling constant $g_X$ as a function of momentum transfer $Q$. The solid blue and gold curves correspond to the boundary condition $\epsilon_\textrm{IR} = 0$ and $\epsilon_\textrm{UV} = 0$, respectively. The vertical, dashed black lines indicate the values of $Q$ for which we match the leptons masses and the shaded red (orange) region represents the WIMP detection window in {\sc Lz} ({\sc Xenon}n{\sc T}), for DM masses between 10 and 1000 GeV.}
    \label{fig:kinmix}
\end{figure}

For very small momentum transfer values $Q\ll m_{i,j}$, the loop–induced kinetic mixing saturates to a constant that is labeled as the infrared limit $\epsilon_{\rm{IR}}$:  
\begin{equation}
\epsilon_{\text{IR}}
     \;=\;
     \epsilon
     \;-\;
     \frac{e\,g_X}{12\pi^{2}}
     \log\frac{m_j^{2}}{m_i^{2}}.
\end{equation} 
Instead, in the ultraviolet limit $Q\gg m_{i,j}$ the loop induced kinetic mixing becomes zero and the form factor tends to the tree-level kinetic mixing $\epsilon_{\text{UV}} = \epsilon$.
Consequently:
\begin{equation}
  \epsilon_{\text{UV}}-\epsilon_{\text{IR}}
  \;=\;
  \frac{e\,g_{X}}{12\pi^{2}}
  \log\frac{m_{j}^{2}}{m_{i}^{2}}.
  \label{eq:epsDiff}
\end{equation}
Expanding the full evaluation of \(\epsilon_{\text{TOT}}(Q)\) around the IR and UV regimes one finds:
\begin{align}
  \epsilon_\text{TOT}(Q)
  &\simeq
  \epsilon_{\text{IR}}
  \;+\;
  \frac{e\,g_{X}}{60\pi^{2}}\,
  \frac{Q^{2}}{m_{i}^{2}},
  &&  (Q \ll m_{i}),
  \label{eq:epsLowQ}
  \\[6pt]
  \epsilon_\text{TOT}(Q)
  &\simeq
  \epsilon_{\text{UV}}
  \;-\;
  \frac{e\,g_{X}}{2\pi^{2}}\,
  \frac{m_{j}^{2}}{Q^{2}},
  &&  (Q \gg m_{j}),
  \label{eq:epsHighQ}
\end{align}
if $m_i<m_j$.
Equations~\eqref{eq:epsDiff}–\eqref{eq:epsHighQ} reproduce the well-known
interpolation between the infrared plateau and the ultraviolet constant,
with the leading momentum-dependent corrections shown explicitly.
We show in Fig.~\ref{fig:kinmix} the shape of $\epsilon_{\rm{TOT}}(Q)$ in the $\epsilon_{\rm{UV}}=0$ and $\epsilon_{\rm{IR}}=0$ regimes.
In particular, we can see that the minimal kinetic mixing parameter is obtained in the limit $Q \ll m_i$ for $\epsilon_{\rm{IR}}=0$, and $Q \gg m_j$ for $\epsilon_{\rm{UV}}=0$, which has $\epsilon = 0$.
However, in the typical range of momentum transfer values of a WIMP DM particles $m_{\chi}\in[10,1000]$ GeV, which is $Q\in[2,400]$ MeV/c, the value of $|\epsilon_\text{TOT}(Q)/g_X|$ is typically smaller than 0.04.


\begin{figure}
\centering
\includegraphics[width=0.99\linewidth]{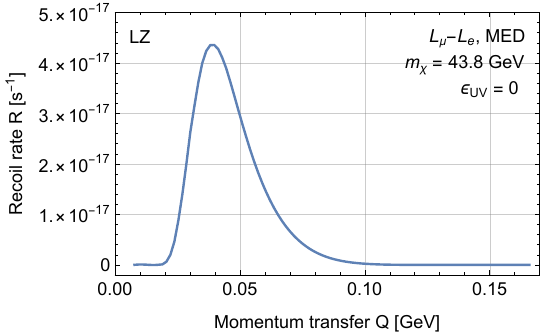}
\includegraphics[width=0.99\linewidth]{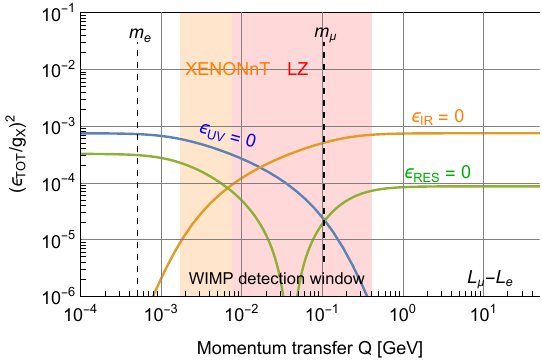}
    \caption{Upper panel: Nuclear recoil rate $R$ at {\sc Lz} in terms of the transfer momentum $Q$ for the $L_\mu-L_e$ model in the case $\epsilon_\text{UV}=0$. Lower panel: Variation of $(\epsilon_{\rm{TOT}}/g_X)^2$ in the $L_\mu-L_e$, assuming three cases for the tree-level component: $\epsilon_\text{UV}=0$ (blue), $\epsilon_\text{IR}=0$ (orange), and $\epsilon_\text{RES}=0$ (green).}
    \label{fig:kinmix2}
\end{figure}

There is a way to suppress the total kinetic mixing for direct detection by taking $\epsilon$ with about the same value of the loop induced one at the momentum transfer relevant for direct detection:
\begin{eqnarray}
\epsilon_\text{RES} \! &=& \! \epsilon_{\rm{TOT}}(Q_\text{RES}) = 0,
\end{eqnarray}
where $Q_\text{RES}$ is the momentum transfer that maximises the nuclear recoil spectrum. For example in the top panel of Fig~\ref{fig:kinmix2} we show the recoil rate distribution at {\sc Lz} as a function of $Q$ for the $L_\mu-L_e$ model when fixing $m_\chi=43.8$ GeV (which will be relevant in Sec.~\ref{sec:results}). The peak of the recoil rate is achieved for $Q\approx 0.04$ GeV.
Therefore, following this approach, the tree-level kinetic mixing is fixed as follows in the $L_\mu-L_e$ model:
\begin{eqnarray}
\label{eq:epsilonres}
\epsilon \! &=& \! \frac{e \, g_{X}}{2\pi^2} \! \int_0^1 \! dx x(1 - x) \! \log \! \left( \frac{m_{e}^2 + Q^2 x(1 - x)}{m_{\mu}^2 + Q^2 x(1 - x)} \right) \!\Bigg|_{Q=0.04\,{\rm{GeV}}} \\
&\sim& 0.7 \times \frac{e \, g_{X}}{2\pi^2} = 1.8 \times 10^{-4}. \nonumber
\end{eqnarray}
In the range of $Q$ that is the most relevant for direct detection, we obtain a suppression even by a factor of 10 with respect to the IR and UV cases reported above  (see lower panel of Fig.~\ref{fig:kinmix2}). Moreover, the value of $\epsilon$ that we need is perfectly compatible with the upper limit on the kinetic mixing obtained from collider experiments, which dictate $\epsilon\leq 10^{-3}$ (see, \textit{e.g.}, \cite{LHCb:2019vmc}).

In the rest we will consider the following cases:
\begin{itemize}
    
\item $\epsilon_\text{UV}=0$, which is equivalent to take zero tree-level kinetic mixing $\epsilon=0$, 
\item $\epsilon_\text{IR}=0$, for which the tree-level kinetic mixing $\epsilon = \frac{e\,g_X}{12\pi^{2}} \log\!\frac{m_{j}^{2}}{m_{i}^{2}}$,
\item $\epsilon_{\rm{RES}}$ with $\epsilon$ evaluated as reported in Eq.~\ref{eq:epsilonres}.
\end{itemize}
We remind the reader that in all the three cases we evaluate the loop induced component of the kinetic mixing by taking into account the full dependence with the momentum transfer as reported in Eq.~\ref{eq:epsilontot}.

The kinetic mixing in the $B-L$ model is not so relevant for the study of its phenomenology because all SM fermions are charged under $U(1)_{B-L}$ thus the vertexes with $A'_\mu$ does not require the kinetic mixing term, \textit{i.e.}~they are dominated by the term proportional to $g_X$.

\section{Model constraints and potential observables}
\label{sec:observables}

\subsection{Magnetic moment}

Precision measurements of the electron and muon anomalous magnetic moments,
\begin{equation}
a_\ell \;\equiv\;\frac{g_\ell-2}{2}\qquad(\ell=e,\mu),
\end{equation}
provide stringent probes of any new $U(1)$ gauge boson
($A'$) that couples to charged leptons.
The most recent values for the difference of the theoretical and experimental values of $a_\ell$ are the following:
\begin{eqnarray}
\label{eq:a-excess}
\Delta a_\mu &=& a_\mu^{\mathrm{exp}}-a_\mu^{\mathrm{SM}}
          = (38\pm\,63)\times10^{-11} \\   
\Delta a_e &=& a_e^{\mathrm{exp}}-a_e^{\mathrm{SM}} = (-\,1.44\pm0.72)\times10^{-12}
\end{eqnarray}
The value reported for $\Delta a_{\mu}$ comes from the difference between the latest theoretical estimate for the SM provided in Ref.~\cite{Aliberti:2025beg} and the latest measurements made at Fermilab \cite{Muong-2:2023cdq}. Instead, $\Delta a_{e}$ is obtained from the experimental data in Ref.~\cite{Fan:2022eto} and the theoretical evaluation performed in Ref.~\cite{Aoyama:2017uqe}.

For a new vector boson $A'$ with coupling $g_X$ and mass $m_{A'}$ the one-loop shift to $a_\ell$ is obtained as (see \textit{e.g.}, Ref.~\cite{Holdom:1985ag}):
\begin{equation}
\Delta a_\ell^{A'} \;=\;
\frac{(g_X + e \epsilon)^2}{4\pi^{2}}\int_0^1\!dz\;
\frac{m_\ell^{2}\,z^{2}(1-z)}{m_{A'}^{2}(1-z)+m_\ell^{2}z^{2}}.
\label{eq:dala}
\end{equation}
Therefore, both the tree level and the loop induced kinetic mixing can contribute to the magnetic anomaly. 
In the case $\epsilon_{\rm{UV}}=0$ the contribution of the kinetic mixing is exactly zero, instead for $\epsilon_{\rm{IR}}=0$ and $\epsilon_{\rm{RES}}=0$ the kinetic mixing contribution to the magnetic moment anomaly can be relevant. 
In the limit $m_{A'}\!\gg\!m_\ell$, $\Delta a_\ell^{A'}$ reduces to
\(\Delta a_\ell^{A'} \simeq g_X^2\,m_\ell^{2}/(12\pi^{2}m_{A'}^{2})\).


\textbf{$\bm{U(1)_{L_\mu-L_{\tau}}}$.}  The model $U(1)_{L_\mu-L_\tau}$ model receives constraints only from $\Delta a_{\mu}$ through the parameter combination \(g_X/m_{A'}\).
In particular, if we require that we do not overshot the $2\sigma$ upper limit for $\Delta a_{\mu}$ we obtain:
  \begin{equation}
    g_X \lesssim 4.2 \times10^{-3}
    \left(\frac{m_{A'}}{\mathrm{1\;GeV}}\right),
    \,
    m_{A'} \geq \mathcal{O}(1)\;\mathrm{GeV}.
    \label{eq:Lmutau-band}
  \end{equation}

\textbf{$\bm{U(1)_{L_e-L_\tau}}$.}  
In this model the muon is neutral under $U(1)_{L_e-L_\tau}$, so $(g-2)_\mu$ \emph{does not} constrain the coupling parameters.  The electron, however, carries charge \(+1\), and the $2\sigma$ upper limit of $\Delta a_e$ forces the couplings to satisfy the following relation:
\begin{equation}
     g_X \lesssim 2.6\times10^{-2}\left(\frac{m_{A'}}{\mathrm{1\;GeV}}\right), \,
    m_{A'} \geq \mathcal{O}(1)\;\mathrm{GeV}
\end{equation}

\textbf{$\bm{U(1)_{L_\mu-L_e}}$.}  
  Here both the muon \emph{and} the electron receive opposite-sign
  charges \((q_{X,\mu}=+1,\;q_{X,e}=-1)\).
  This model can contribute to both $\Delta a_\mu$ and $\Delta a_{e}$ but, as seen before, the constraints coming from $\Delta a_\mu$ are much tighter. 
Imposing the bound $\Delta a_\mu$ yields to an upper limit for $g_X$ which is the same of the $L_{\mu}-L_{\tau}$ model.

\textbf{$\bm{U(1)_{B-L}}$}. The upper limits on $g_X$ obtained for this model are dominated by the constraints on the muon magnetic moment and are the same as in Eq.~\ref{eq:Lmutau-band}.

\subsection{Relic density}
\label{sec:relic}

The present-day DM relic density has been precisely measured by the Planck satellite, yielding $\Omega_\chi h^2 = 0.120 \pm 0.001$ with a relative uncertainty at the level of 1\%~\cite{Planck:2018vyg}. 

The theoretical prediction of the DM relic density requires solving the Boltzmann equation that governs the phase-space distribution $f_\chi(\vec{p})$ of the DM particle $\chi$ in an expanding Friedmann–Robertson–Lemaître–Walker Universe~\cite{Kolb:1990vq,Edsjo:1997bg}:
\begin{equation}
E \left( \partial_t - H \vec{p} \cdot \nabla_{\vec{p}} \right) f_\chi(\vec{p}) = \mathcal{C}[f_\chi]\,,
\label{eq:RD1}
\end{equation}
where $E$ and $\vec{p}$ are the energy and momentum of the DM particle, and $H$ is the Hubble expansion rate. The right-hand side, $\mathcal{C}[f_\chi]$, is the collision operator that encodes all relevant interactions between DM and SM particles. It contains contributions from elastic scattering ($\mathcal{C}_{\rm el}$), which maintains kinetic equilibrium, and from annihilation processes ($\mathcal{C}_{\rm ann}$), which control chemical equilibrium. For a detailed formulation of both components, see Ref.~\cite{Binder:2017rgn}.

In the standard thermal freeze-out scenario of WIMPs, several assumptions are commonly made to simplify the problem. Most notably, one assumes that kinetic equilibrium between DM and the SM bath persists throughout chemical decoupling. Under this condition, the phase-space distribution of $\chi$ remains proportional to the thermal equilibrium distribution, \textit{i.e.}, $f_\chi \propto f_{\chi,\text{eq}}$. This approximation reduces Eq.~\eqref{eq:RD1}—a partial differential equation—to a simpler ordinary differential equation describing the evolution of the DM number density, which is called the Lee-Weinberg equation:
\begin{equation}
\frac{\mathrm{d} n_\chi}{\mathrm{d} t} + 3 H n_\chi = -\left\langle \sigma v_\textrm{M\o l} \right\rangle_T \left(n_\chi^2 - n_{\chi, \mathrm{eq}}^2\right)\,,
\label{eq:RD2}
\end{equation}
where $n_\chi = g_\chi \int \frac{d^3 p}{(2\pi)^3} f_\chi(p)$ is the $\chi$ number density, and $\left\langle \sigma v_\textrm{M\o l} \right\rangle_T$ is the thermally averaged annihilation cross section evaluate at temperature $T$ in the early Universe.

In the non-relativistic regime, where $f_{\chi,\text{eq}} \simeq \exp(-E/T)$, the thermal average of the annihilation cross section takes the standard form \cite{GONDOLO1991145}:
\begin{equation}
\left\langle \sigma v_\textrm{M\o l} \right\rangle_T = \int_{4 m_\chi^2}^{\infty} \mathrm{d} s \, \frac{s \sqrt{s - 4 m_\chi^2} \, K_1(\sqrt{s}/T) \, \sigma v_{\rm M\o l}(s)}{16 T m_\chi^4 K_2^2(m_\chi/T)}\,,
\end{equation}
where $K_i$ are the modified Bessel functions of the second kind of order $i$.

This framework provides the basis for accurately computing the DM relic density for a wide range of particle physics models under the WIMP paradigm.

In the \( U(1)_{L_\mu - L_\tau} \) extension of the DM, and similarly for the other two models, the dominant annihilation channels are mediated by the new gauge boson \( A'_\mu \), which couples to \( \chi \) with strength \( g_X q_X \) and to the muon and tau sectors via the coupling $g_X$ and their \( L_\mu - L_\tau \) charges $q_{X,f}=\pm1$. 

In the $B-L$ model, DM particles annihilate into quarks and leptons without kinetic mixing suppression. 
In particular, the branching ratio for annihilation into each lepton pair is a factor of 3 larger than that into each quark pair. 
The reason is that the quark $B-L$ charge is a factor $1/3$ smaller than that of leptons (giving an overall factor $1/9$ in the cross section), although this is partially compensated by summing over the three quark colors.

Considering the $L_{\mu}-L_{\tau}$ as an example, the primary annihilation modes contributing to thermal freeze-out are:
\begin{align}
\chi \bar{\chi} \rightarrow A'^* \rightarrow \mu^+ \mu^-,\; \tau^+ \tau^-,\; \nu_\mu \bar{\nu}_\mu,\; \nu_\tau \bar{\nu}_\tau,
\end{align}
for which the cross section is evaluated as:
\begin{equation}
\sigma(s) = \sum_f \frac{k_f\, g_X^4 q_X^2}{12\pi s} \cdot \frac{\beta_f}{\beta_\chi}
\left[
\frac{(s + 2m_\chi^2)(s + 2m_f^2)}{(s - m_{A'}^2)^2 + m_{A'}^2 \Gamma_{A'}^2}
\right],
\end{equation}
where $s$ is the Mandelstam variable, $k_f$ is equal to 1 for $e,\mu,\tau,\chi$ and $1/2$ for neutrinos, $\beta_i = \sqrt{1 - 4m_i^2 / s}$, and $\Gamma_{A'} = \Gamma_{A' \to \chi\bar{\chi}} + \sum_f \Gamma_{A' \to f\bar{f}}$,
with the sum over \( f = \mu,\, \tau,\, \nu_\mu,\, \nu_\tau \).
In case of the $B-L$ model the cross section should be multiplied for the number of colors where DM annihilates into quarks.
The decay width of the $A'$ decaying into $f\bar{f}$ is given by:
\begin{equation}
\Gamma_{A' \to f\bar{f}} = k_f m_{A'} \times \frac{g_X^2 q_{X,f}^2}{12\pi} 
\sqrt{1 - \frac{4m_f^2}{m_{A'}^2}} 
\left( 1 + \frac{2 m_f^2}{m_{A'}^2} \right).
\end{equation}
This equation is also valid for the $A'$ that decays into DM pairs with the simple substitution of $q_{X,f}$ into $q_X$.
For the decay of the $A'$ into quarks or the lepton neutral for the model one has to replace $g_X q_X$ with $\epsilon(Q^2)e$.

In order to determine an approximated result for the thermally averaged cross section, we can  expand the center of mass energy in relative velocity $v$ of the DM annihilating particles
\(s \simeq 4 m_\chi^{2} (1 + v^{2}/4)\).
Retaining only the $s$-wave contribution (appropriate for a vector mediator and Dirac dark matter) one obtains
\begin{multline}
\langle \sigma v \rangle \simeq
\sum_{f}
\frac{q^2_Xg_X^{4} k_f}{2\pi}\,
\frac{m_\chi^{2}}
{\left(m_{A'}^{2}-4m_\chi^{2}\right)^{2}+m_{A'}^{2}\Gamma_{A'}^{2}} \times \\
\times
\sqrt{1-\frac{m_f^{2}}{m_\chi^{2}}}
\left(1+\frac{m_f^{2}}{2m_\chi^{2}}\right)
+\mathcal{O}(v^{2}),
\label{eq:sigmav_swave}
\end{multline}
where the sum runs over
\(f=\mu,\,\tau,\,\nu_\mu,\,\nu_\tau\). The zeroth–order term in Eq.~\eqref{eq:sigmav_swave} controls the standard \emph{$s$-wave} relic abundance. Near the resonance region \(m_{A'}\simeq 2m_\chi\), the Breit–Wigner denominator enhances the cross section, allowing smaller values of \(g_X q_X\).


After chemical decoupling the comoving number density $Y_0 \equiv n_\chi/s$ is
approximately\footnote{
  We follow the standard treatment of
  Refs.~\cite{Kolb:1990vq,GONDOLO1991145}.  The freeze-out variable is
  $x_f \equiv m_\chi/T_f \simeq 20$--$30$ for a weak‐scale WIMP; 
  $g_\ast \simeq 90$ counts the relativistic degrees of freedom at
  $T_f$.}:
\begin{equation}
Y_0 \;\simeq\;
\frac{\,3.79\;x_f\,}{\,\sqrt{g_\ast}\;M_{\rm Pl}\,m_\chi\,
      \langle \sigma v\rangle}\; .
\end{equation}
The present–day relic abundance then reads:
\begin{equation}
\Omega_\chi h^{2}
\simeq
\frac{m_\chi s_0 Y_0}{\rho_{\rm c}}
\;\approx\;
1.07 \times 10^{9}\;{\rm GeV}^{-1}\;
\frac{x_f}{\sqrt{g_\ast}\,M_{\rm Pl}}\;
\frac{1}{\langle \sigma v\rangle}
\label{eq:omega_generic}
\end{equation}
with Plank mass $M_{\rm Pl}=1.22\times10^{19}\,{\rm GeV}$, critical density
$\rho_{\rm c}=8.05\times10^{-47}\,h^2\;{\rm GeV}^4$, and today's entropy density
$s_0 = 2.89\times10^3\;{\rm cm}^{-3}$.
Demanding $\Omega_\chi h^{2}\simeq 0.12$ fixes the annihilation cross section to the following value
\begin{equation}
\langle\sigma v\rangle \;\simeq\;
3\times10^{-26}\;{\rm cm^3\,s^{-1}}
\left(\frac{x_f}{25}\right)\!
\left(\frac{90}{g_\ast}\right)^{\!\!1/2},
\label{eq:sigmav_needed}
\end{equation}
which for $x_f\sim 25$ and $g_{\ast}\sim 90$ provides the typical value for the thermal cross section needed in the standard WIMP scenario to obtain the correct DM relic abundance.

In our models we consider Dirac DM charged under $U(1)_{L_i-L_j}$ or $U(1)_{B-L}$ for which the $s$–wave annihilation cross section is give in Eq.~\ref{eq:sigmav_swave}.
We can take two extreme cases below and above the resonance region ($ m_{A'} \sim 2 m_\chi$).

\begin{itemize}

\item Let's start with the case above the resonance for which $m_{A'} \gg 2 m_{\chi}$. We also require that $m_\chi\gg m_f$. The values of $g_X q_X$ that match the relic density are given by
\begin{equation}
g_X q_X  \simeq 
8.0 \times 10^{-3}\,
\left(\frac{x_f}{25}\right)^{1/4} \left(\frac{g_\ast}{90}\right)^{-1/8}\!
\frac{\bigl(m_{A'}/\text{GeV}\bigr)}
     {\bigl(m_\chi/\text{GeV}\bigr)^{1/2}} \ .
\end{equation}
Therefore, for the typical value of $x_f$ and $g_{\ast}$ of the freeze-out process, $m_\chi$ of a few tens of GeV that fit the GCE spectrum, and $A'$ boson mass of 100 GeV the coupling values that match the relic density is $g_X q_X  \simeq 0.1-0.3$ and increases linearly with $m_{A'}$.

\item Instead, in the opposite regime below the resonance for which $ m_{A'} \ll 2 m_\chi $ the values of $g_X q_X$ that matches the relic density are independent by $m_{A'}$ and given by
\begin{equation}
\label{eq:relicabove}
g_X q_X  \simeq 
1.4 \times 10^{-2}\,
\left(\frac{x_f}{25}\right)^{1/4} \left(\frac{g_{\ast}}{90}\right)^{1/4}
\left(\frac{m_\chi}{\text{GeV}}\right)^{1/2} \ .
\end{equation}
Choosing $m_\chi$ of a few tens of GeV as before, the coupling values that match the relic density is $g_X q_X  \simeq 0.04-0.1$.

\item In order to make an example for the region around the resonance one we can assume that the DM mass is $2m_\chi=m_A(1+\xi)$ and $m_\chi\gg m_f$. Therefore, we take the mass of $A'_\mu$ a factor $\xi$ away from the resonant peak.
In this case, the values of $g_X q_X$ that match the relic density are:
\begin{equation}
g_X q_X \!=\! 1.8 \times 10^{-2} \!
\left(\frac{x_f}{25}\right)^{1/4} \!
\left(\frac{90}{g_\ast}\right)^{1/8} \!
\left(\frac{m_{A'}}{\text{GeV}}\right)^{1/2}\!\xi^{1/2}.
\end{equation}
For $\xi=0.1$ and $m_A'$ in the range between 10 and 80 GeV we get $g_X q_X\sim 0.01-0.03$, which is smaller with respect to the previous two cases. Moreover, the smaller is the value of $\xi$, \textit{i.e.}~the closer we are to the resonance region, and the smaller will be the values of $g_X q_X$ needed to get the right relic density.
Therefore, the advantage of considering the resonant region is that we can achieve the relic density for values of the couplings much smaller than the previous two cases.
\end{itemize}

In order to compute the parameter space of the $L_i-L_j$ and $B-L$ models that matches the DM relic density, we use {\tt micrOMEGAs}~\cite{Alguero:2023zol}, which solves numerically Eq.~\ref{eq:RD2}. The {\tt CalcHEP}~\cite{Belyaev:2012qa} model files we need to provide {\tt micrOMEGAs} for the computations are generated using {\tt FeynRules}~\cite{Alloul:2013bka}.

\subsection{Direct detection}
\label{sec:direct}

\subsubsection{Electron scattering}

Kinetic mixing between the dark photon $A'_{\mu}$ and the SM photon induces a
non–zero elastic DM–electron scattering cross section
(see, \textit{e.g.}~Ref.~\cite{Essig:2015cda}):
\begin{equation}
\sigma_{e}(Q)
= \frac{\mu_{\chi e}^{2}e^2q_X^2g_X^2\epsilon_\text{TOT}^{2}(Q)}
       {\pi\bigl(m_{A'}^{2}+Q^{2}\bigr)^{2}},
\label{eq:sigma_electron}
\end{equation}
where $\mu_{\chi e}$ is the DM–electron reduced mass.
Experiments such as {\sc Xenon}n{\sc T}~\cite{XENON:2024znc} recently released new upper limits for the DM–electron scattering cross sections but they are sensitive to DM masses below 1 GeV, which is not the focus of this paper.
Therefore, we will not consider the constraints coming from electron-recoil scattering.
Moreover, for $\epsilon_\text{UV}=0$ and independently on the $U(1)_{L_i-L_j}$ model considered, we find that for $m_\chi \in [0.06, 1]$ GeV, the limit on $g_Xq_X$ is of the order $g_Xq_X \lesssim (m_{A'}/1\,\textrm{GeV})$, assuming a fixed momentum transfer of $Q = \alpha m_e$.

\subsubsection{Nuclear scattering}

Analogously, kinetic mixing generates elastic DM–nucleon scattering with a spin–independent cross section off a nucleus $N$~\cite{Evans:2017kti}:
\begin{multline}
\sigma_{N}(Q)
= \frac{1}{\pi}\,\mu_{\chi N}^{2} q_X^2g_X^{2}\,\epsilon_{\rm{TOT}}^{2}(Q)F^2_\text{Helm}(Q)
  \times \\ 
  \times
  \left|
    \frac{f_{N}^{(A')}}{m_{A'}^{2}+Q^2} - 
    \frac{s_W f_{N}^{(Z)}}{(m_{Z}^{2}-m_{A'}^{2})}\right|^{2},
\label{eq:sigma_nuclear}
\end{multline}
with $F_\text{Helm}(Q)$ is the Helm nuclear form factor~\cite{Helm:1956zz,Duda:2006uk}, $\mu_{\chi N}$ the DM–nucleon reduced mass and
\begin{equation}
f_{N}^{(X)} =
\frac{1}{A}\left[
      Z\bigl(2g_{uX}+g_{dX}\bigr)
      + (A-Z)\bigl(g_{uX}+2g_{dX}\bigr)
\right],
\label{eq:fN_def}
\end{equation}
where $X = A',\,Z$ denotes the relevant gauge bosons involved in the DM-nucleon scattering, while $A$ and $Z$ denote the atomic mass and charge of the target nucleus. 
The factors $g_{qX}$ are the chiral couplings of quark
$q=u,d$ to the boson $X$, as expressed in Ref.~\cite{Arcadi:2018tly}.
The strongest current limit on $\sigma_{N}$ arises from the {\sc Lz} experiment \cite{LZ:2022ufs} for DM masses above $\sim 8.9$ GeV. Instead, for $6\, \text{GeV} \lesssim m_\chi \lesssim 8.9\, \text{GeV}$ {\sc Xenon}n{\sc T} provides the strongest limits~\cite{XENON:2023cxc}.

The experimental limits on $\sigma_N$ are usually provided with the zero momentum transfer assumption. Therefore, in order to set a limit on $g_X$ that takes into account the momentum transfer dependency in $\sigma_N$, we have to compute the recoil rate directly, and compare it to the experimental one. The recoil rate is written as:
\begin{equation}
\label{eq:rate}
    R = \frac{A^2}{2 \mu^2_{\chi N}}\left(\frac{\rho_\odot}{m_\chi}\right)\int_{E^\text{min}_R}^{E^\text{max}_R} \sigma_N(Q(E_R))\; \omega(E_R)\;\eta(E_R)\;dE_R\;,
\end{equation}
where we integrate over the recoil energy $E_R = Q^2/2M_T$, with $M_T$ being the mass of the target nucleus. $A$ is the mass number of the target nucleus\footnote{In {\sc Lz} and {\sc Xenon}n{\sc T}, which use Xe targets, we have $A\sim131$.}, $\rho_\odot$ is the local DM density, $\omega(E_R)$ is the detector efficiency and $E_R^\text{min}$ is the detection threshold energy, which we take from Ref.~\cite{LZ:2022ufs} for {\sc Lz} and Ref.~\cite{XENON:2023cxc} for {\sc Xenon}n{\sc T}. $E^\text{max}_R = 2\mu^2_{\chi T}(v_\text{esc}+v_\odot)^2/M_T$ is the maximal recoil energy, for a DM particle with a velocity equal to the escape velocity $v_\text{esc} \simeq 550$ km/s, that would free this particle from the gravitational potential of the Milky Way. Here $\mu_{\chi T}$ denotes the reduced mass of the DM-target system. Finally, $\eta(E_R)$ is related to the velocity distribution $f(\vec{v})$ of DM particles in the Milky Way:
\begin{equation}
    \eta(E_R) = \int_{v_\text{min}}^{v_\text{esc}} \frac{f(\vec{v})}{|\vec{v}|}\,d^3\vec{v} \;,
\end{equation}
where we assume that $f(\vec{v}) \propto \exp(-|\vec{v}+\vec{v}_\odot|^2/v_0^2)\,\Theta(v_\text{esc}-|\vec{v}+\vec{v}_\odot|)$ is a truncated Gaussian function (normalised so that $\int_0^{v_\text{esc}} f(\vec{v}) \,d^3\vec{v}=1$), with standard deviation $v_0 \simeq 235$ km/s and $\Theta$ is the Heaviside function. We perform a Gallilean transformation so the DM particle velocity is $\vec{v}+\vec{v}_\odot$ in the laboratory frame\footnote{Choosing $\vec{x}$ as the GC-Sun axis, $\vec{v}_\odot$ has only a component in $\vec{y}$. Therefore, in spherical coordinates, $\vec{v}+\vec{v}_\odot = (v\sin\theta\cos\phi,v\sin\theta\sin\phi+v_\odot,v\cos\theta)$.}, where $|\vec{v}_\odot| \simeq 220$ km/s is the velocity of the Sun in the Galactic frame.

Specifically, we compute on one hand the experimental upper limit for the recoil rate $R$. This is done by using Eq.~\ref{eq:rate}, fixing $Q=0$ and taking for $\sigma_N$ the published upper limits.
Then, we compare the experimental upper limit for $R$ with the recoil rate predicted by our model, evaluated taking into account the momentum dependence in Eq.~\ref{eq:rate}, in order to set an upper limit on $g_X q_X$.

As reported in the previous section, for DM masses well below resonance region ($m_{A'}\ll 2 m_{\chi}$), the relic density does not depend on the mediator mass. For example for the $U(1)_{L_{\mu}-L_{e}}$ model, $q_X=1$, and DM mass of $50$ GeV, the correct relic density is reached at a value of the order of $g_X \sim 9 \times 10^{-2}$ (see Eq.~\ref{eq:relicabove}).
In the range of transverse momentum of direct detection and WIMP particles, $|\epsilon_{\rm{TOT}}(Q)| \leq 0.02 \cdot g_X$ (see the bottom panel of Fig~\ref{fig:kinmix}). This implies that considering the $g_X$ needed to fit the relic density we obtain a total kinetic mixing factor of the order of $\epsilon \leq 1.6 \times 10^{-3}$. Using $\epsilon = 1.6 \times 10^{-3}$ and $Q=0.04$ GeV (the peak momentum transfer at {\sc Lz}) into the Eq.~\ref{eq:sigma_nuclear}, we obtain a value of the cross section that can reach as low as $\sigma_N \approx 1.3\times 10^{-44} \,\textrm{cm}^2\,(m_{A'}/1\,\textrm{GeV})^4$ (assuming DM to be a thermal relic), which is about $4$ orders of magnitude larger than the {\sc Lz} upper limit for $m_{A'}=1$ GeV.
Therefore, in order to make the model compatible with the direct detection upper limits we should either consider a scenario that suppresses even more the kinetic mixing, for example the $\epsilon_{{\rm{RES}}}$ case or choose DM and mediator masses closer to the resonance regime for which we can obtain the relic density with smaller values of $g_X$. 

The model $B-L$ does not suffer from a suppressed kinetic mixing interaction of DM with quarks. Therefore, most of the parameter space is ruled out by nuclear direct detection upper limits. 
The theoretical DM-nucleon scattering cross section is given by the following equation~\cite{Arcadi:2023lwc}:
\begin{equation}
    \sigma_N(Q) = \frac{\mu_{\chi N}^2q_X^2g_X^4}{\pi (m_{A'}^2+Q^2)^2}F^2_\text{Helm}(Q)
\end{equation}
For example if the DM particle has a mass of 50 GeV and fixing $g_X$ to be $5\times10^{-2}$ (for $m_{A'}\ll 2m_\chi$) to match the right relic density, the direct detection upper limit is $\sigma_N \approx 6\times10^{-34}\,\textrm{cm}^2\,(m_{A'}/1\,\text{GeV})^4$, \textit{i.e.} around 9 orders of magnitude larger than in $L_i-L_j$ models. This implies that the upper limits on $g_X$ are expected to be a factor of about 100 stronger.

\subsection{Collider constraints}
\label{sec:collider}

Collider experiments such as {\sc BaBar}, {\sc Na64}-e, {\sc Belle}~II, {\sc Atlas}, and {\sc Cms} have searched for the decay of the dark photon $A'_\mu$ into fermions.
For example, {\sc BaBar} and {\sc Belle}~II looked for the process $e^{+}e^{-}\!\to\! Z'\,\mu^{+}\mu^{-}$~\cite{BaBar:2016sci,Belle-II:2022yaw} and $e^{+}e^{-}\!\to\! Z'\,\gamma$~\cite{BaBar:2014zli} followed by $Z'\!\to\!\mu^{+}\mu^{-}$.
No excess over the expected SM background was found, and upper limits were set on the cross section, the kinetic mixing ($\epsilon \lesssim 5\times10^{-3}$ for $m_{A'} < 10$ GeV) as well as on the gauge coupling $g_{X}$, ranging from $g_{X}\lesssim(1\!-\!4)\times10^{-3}$ at $m_{A'}=1~\text{GeV}$ to $g_{X}\lesssim0.1$ at $m_{A'}=8~\text{GeV}/c^{2}$.

The limits obtained by the {\sc Na64}-e Collaboration are comparable to those of {\sc BaBar} and {\sc Belle}~II, but were derived by impinging 100~GeV electrons on an active thick target~\cite{NA64:2022rme}.

The {\sc Cms} and {\sc Atlas} experiments searched for dark photons in $pp$ collisions at $\sqrt{s}=13~\text{TeV}$ via the process $pp\to Z\to4\mu$~\cite{ATLAS:2023vxg,CMS:2018yxg}.
These searches are sensitive to $A'$ masses up to about $80~\text{GeV}$.

\medskip

Muon neutrinos scattering off the Coulomb field of a nucleus can produce a pair of muons ($\nu_{\mu}N\to\nu_{\mu}\mu^{+}\mu^{-}N$).
This process, called \emph{neutrino trident production}, happens in the SM through the exchange of a $W$ or a $Z$ boson.
The {\sc Ccfr} Collaboration~\cite{CCFR:1991lpl} measured the ratio of the observed trident cross section to the SM prediction to be
$\sigma_{\text{{\sc Ccfr}}}/\sigma_{\text{SM}}=0.82\pm0.28$.
Although statistically consistent with the SM, the central value below unity leaves room for contributions from physics beyond the SM.
Ref.~\cite{Altmannshofer:2014pba} showed that trident production can also proceed via a new vector boson $A'$, enabling constraints on the coupling $g_X$.
For $4~\text{GeV}\le m_{A'}\le6~\text{GeV}$ this remains the most stringent bound to date.
It is worth stressing that only muon neutrinos yield an experimentally accessible $\mu^{+}\mu^{-}$ trident rate: charged--current diagrams are present for $\nu_{\mu}$ but absent for $\nu_{e}$ and $\nu_{\tau}$, which therefore contribute only through suppressed neutral--current channels.

\medskip

Finally, the {\sc Lhc}b experiment has searched for prompt and long-lived dark photons produced in proton--proton collisions at $\sqrt{s}=13~\text{TeV}$ and decaying into two muons~\cite{Ilten:2016tkc}.
The resulting limits on the kinetic mixing parameter cover $m_{A'}\in[0.1,80]~\text{GeV}$ and range from $10^{-4}$ to a few~$\times10^{-3}$ for $g_{X}=0.1$.

\subsection{{\sc Fermi-Lat} Galactic Centre excess}

The excess of $\gamma$-rays detected in \textsc{Fermi-Lat} data is one of the most promising potential DM signatures. Nevertheless, a population of millisecond pulsars in the Galactic bulge remains a perfectly viable interpretation \cite{Macias:2016nev,Bartels:2017vsx,Manconi:2024tgh}.

The energy-flux spectrum of the GCE peaks at a few GeV, as expected for DM particles annihilating into hadronic final states—\textit{i.e.}, quarks and SM gauge bosons—which produce prompt $\gamma$-rays dominated by hadronization. As already found in previous works (see \textit{e.g.}, Refs.~\cite{Goodenough:2009gk,Hooper:2010mq,Calore:2014nla,Calore:2014xka,DiMauro:2021qcf}), annihilation into quark–antiquark channels with $m_{\chi}\simeq50\,\mathrm{GeV}$ and a cross section close to the thermal value fits the GCE well. Channels involving SM gauge bosons can also reproduce the GCE, but only if the bosons are significantly off shell—typically disfavored because the off-shell branching ratios are small.

When considering only prompt $\gamma$-ray production for leptonic annihilation channels, the predicted DM flux does not fit the GCE \cite{Calore:2014nla,DiMauro:2021qcf}. This changes drastically once secondary processes such as inverse Compton scattering (ICS) and, to a lesser extent, bremsstrahlung from electrons and positrons are included, as already noted in \cite{DiMauro:2021raz}.

In this work we fit the GCE using both the Di Mauro+21~\cite{DiMauro:2021raz} and Cholis+22~\cite{Cholis:2021rpp} datasets. We treat the spread in the best-fit DM mass $m_\chi$ and annihilation cross section $\langle\sigma v\rangle$ obtained from the two fits as an additional source of systematic uncertainty—complementary to the statistical errors obtained when fitt each set individually. For $\langle\sigma v\rangle$, the dominant uncertainty instead arises from the choice of the DM density profile and the associated geometrical $J$-factor, amounting to a factor of $\simeq2$ (see Tab.~\ref{tab:Jfact}).

The two analyses differ mainly in their regions of interest (ROI) and in the interstellar-emission templates adopted. Di Mauro+21 employs a $40^{\circ}\times40^{\circ}$ window centered on the GC, whereas Cholis+22 uses the same nominal ROI but mask the Galactic plane ($|b|<2^{\circ}$) and all resolved point sources. Although the mask removes only about 20\% of the solid angle, it covers the innermost part of the Galaxy, where the DM signal is brightest. As a result, the normalization of the GCE inferred by Cholis+22 is roughly a factor of two lower than that found by Di Mauro+21, as shown in Fig.~\ref{fig:GCEresults}.

\subsubsection{Prompt radiation}
\label{sec:prompt}

The differential prompt flux of $\gamma$-rays from fermionic DM is:
\begin{equation}
\frac{{\rm d}\Phi_\gamma^\text{prompt}}{{\rm d}E_\gamma}
 \;=\;
\frac{r_\odot}{16\pi}\;
\left( \frac{\rho_\odot}{m_{\chi}} \right)^{\!2}\,
\overline{\mathcal J}\;
\langle\sigma v\rangle\,
\sum_f {\rm Br}_f
\left(\frac{{\rm d}N_\gamma}{{\rm d}E_\gamma}\right)_f ,
\label{eq:flux_prompt}
\end{equation}
where $\rho_\odot$ is the DM local density,
$r_\odot=8.122\;\text{kpc}$ the Sun’s Galactocentric distance
\cite{2019ApJ...871..120E},
and $\langle\sigma v\rangle$ the velocity–averaged
annihilation cross section.
The factor $\overline{\mathcal J}$ denotes the line-of-sight
integral, averaged over the solid angle
(typically $\Delta\Omega = 40^{\circ}\!\times40^{\circ}$)
around the GC as in Ref.~\cite{DiMauro:2021raz}.
The sum in Eq.~\ref{eq:flux_prompt} is evaluated over the different annihilation channels.
Channel–dependent photon spectra
$({\rm d}N_\gamma/{\rm d}E_\gamma)_f$ and branching ratios
${\rm Br}_f = \langle\sigma v\rangle_f/\langle\sigma v\rangle$ are derived using {\tt MadDM}~\cite{Backovic:2013dpa,Ambrogi:2018jqj,Arina:2021gfn}, a {\tt MadGraph5\_aMC@NLO}~\cite{Alwall:2014hca, Frederix:2018nkq} plugin for computing DM signals. In particular, for each $U(1)_{L_i-L_j}$ model we have built the the associated UFO~\cite{Degrande:2011ua} files by using {\tt FeynRules} \cite{Alloul:2013bka} and provided them to {\tt MadDM}.
In particular, {\tt MadDM} evaluates the $\langle \sigma v\rangle$ for each annihilation channel and then combined them with the sum in Eq.~\ref{eq:flux_prompt} by using precomputed source spectra from {\tt CosmiXs} \cite{Arina:2023eic,DiMauro:2024kml}.
When $m_\chi>m_{A'}$ the annihilation channel $\chi \bar{\chi} \rightarrow A' A'$ is kinematically open. This channel can contribute to indirect detection because $A'$ decays into leptons. Therefore, in order to include this channel we evaluate the spectra with {\tt MadDM} by evaluating the four-body channel:
\begin{equation}
\chi \bar{\chi} \rightarrow A' A' \rightarrow 2 (\ell_i /\nu_i) \; 2 (\ell_j /\nu_j),
\end{equation}
where $i,j \in \{1,2,3\}$.
See the discussion in Refs.~\cite{DiMauro:2023tho,Arina:2023eic} for more details about the implementation of this process in {\tt MadDM}.


The geometrical factor is defined as:
\begin{equation}
\overline{\mathcal J}
=
\frac{1}{\Delta\Omega}
\int_{\Delta\Omega}\!\!{\rm d}\Omega
\int_{\rm l.o.s.}\!\!\frac{{\rm d}s}{r_\odot}
\left(\frac{\rho(r(s,\Omega))}{\rho_\odot}\right)^2 .
\label{eq:Jbar}
\end{equation}
For $\rho(r)$ we adopt either
a generalised NFW profile,
\[
\rho_{\rm gNFW}(r)
=
\frac{\rho_s}{(r/r_s)^{\gamma}\,[1+r/r_s]^{3-\gamma}},
\]
an Einasto form,
\[
\rho_{\rm Einasto}(r)
=
\rho_s\,
\exp\;\!\left\{-\frac{2}{\alpha}\left[\left(\frac{r}{r_s}\right)^{\alpha}-1\right]\right\}.
\]
Normalisation $\rho_s$ and scale radius $r_s$
can be fixed by fitting the Milky Way rotation curve and the spatial morphology of the GCE.
In particular, rotation curve data constrain the DM density for distances larger than a few kpc from the GC while the GCE spatial morphology fix the inner density profile.
This procedure has been performed in \cite{DiMauro:2021qcf} where they have found for gNFW profile $\gamma\in[1.2-1.3]$ or an for the Einasto profile $\alpha=0.13$.
In addition Ref.~\cite{DiMauro:2021qcf} used the uncertainties of the local DM density profile, which has been measured in Ref.~\cite{2019JCAP...10..037D} to be between $0.3-0.4$ GeV/cm$^3$.
Using the same approach used in Ref.~\cite{DiMauro:2021qcf} and as shown in Tab.~\ref{tab:Jfact}, we have found that the geometrical factor can vary from $120$ to $422$ for the Di Mauro+21 dataset and $67.9$ to $179$ for the Cholis+22 one.
We will use a reference value of the geometrical factor of $209_{-89}^{+213}$ ($89.6_{-21.7}^{+89.4}$) when fitting Di Mauro+21 (Cholis+22) measurement.
The differences in the values of the geometrical factors are due to the fact that the analysis performed in Cholis+22 masks the Galactic plane and point sources therefore the volume integrated is smaller than the one of Di Mauro+21.

\begin{table*}[]
    \centering
    \begin{tabular}{|c|c|c|c|c|c|c|c|}
         \hline
         Label & DM profile & slope ($\gamma/\alpha$) & $\rho_s\,[\text{GeV}/\text{cm}^3]$ & $r_s\,[\text{kpc}]$ & $\rho_\odot\,[\text{GeV}/\text{cm}^3]$ & $\overline{\mathcal{J}}_\text{Di Mauro+21}$ & $\overline{\mathcal{J}}_\text{Cholis+22}$ \\
         \hline
         \verb|MIN| & \multirow{2}{*}{gNFW} & 1.20 & 0.416 & 12.87 & 0.300 & 120 & 67.9 \\
         \cline{1-1} \cline{3-8}
         \verb|MED| &   & 1.30 & 0.449 & 12.67 & 0.345 & 209 & 89.6 \\
         \hline
         \verb|MAX| & Einasto & 0.13 & 0.864 & 5.51 & 0.390 & 422 & 179 \\
         \hline
    \end{tabular}
    \cprotect\caption{Summary of the DM density parameters we use in this work, taken from the \verb|MIN|, \verb|MED| and \verb|MAX| cases in Tab. 1 of Ref.~\cite{DiMauro:2021qcf}. We compute the J-factors of the Di Mauro+21 and Cholis+22 GCE datasets from these parameters.}
    \label{tab:Jfact}
\end{table*}

\subsubsection{Inverse-Compton emission}
\label{sec:ICS}

If DM annihilates preferentially into leptons
($e^+e^-$, $\mu^+\mu^-$, $\tau^+\tau^-$),
the resulting high-energy $e^\pm$ up-scatter the interstellar radiation field (ISRF), yielding a secondary $\gamma$-ray component.
The ISRF energy density near the GC exceeds the local value by roughly an order of magnitude \cite{Porter_2008}, making ICS particularly relevant there.
The differential ICS flux reads
\begin{multline}
\frac{{\rm d}\Phi_\gamma^\text{ICS}}{{\rm d}E_\gamma} =
\frac{r_\odot}{4\pi}
\left(\frac{\rho_\odot}{m_{\chi}}\right)^{2}
\!\int_{\Delta\Omega}{\rm d}\Omega\;
\!\int_{\rm l.o.s.}\frac{{\rm d}s}{r_\odot} \times \\
\times \int_{m_e}^{m_{\chi}}\!\!{\rm d}E_e\;
\mathcal N_e(E_e,\vec{x}(s,\Omega))\;
\mathcal P_\text{ICS}(E_\gamma,E_e,\vec{x}(s,\Omega))
\label{eq:ICS_full}
\end{multline}
where
$\mathcal N_e$ is the propagated $e^\pm$ density and
$\mathcal P_\text{ICS}$ the single-electron ICS power
(see \cite{Cirelli:2010xx,DiMauro:2015tfa}).
As already noted in \cite{DiMauro:2021raz}, in the inner Galaxy energy losses (ICS + synchrotron) dominate over spatial diffusion; we therefore set
$K(E)\to 0$ in the transport equation,
which reduces Eq.~\ref{eq:ICS_full} to
\begin{multline}
\frac{{\rm d}\Phi_\gamma^\text{ICS}}{{\rm d}E_\gamma} \!
= \!
\frac{r_\odot}{{8\pi}}
\left(\frac{\rho_\odot}{m_{\chi}}\right)^{\!2}
\overline{\mathcal J}
\langle\sigma v\rangle
\sum_f
\!\int_{m_e}^{m_{\chi}} {\rm d}E_e \times \\
\times\frac{\mathcal P_\text{ICS}(E_\gamma,E_e)}{b(E_e)}
\mathcal Y_f(E_e)\,
\label{eq:ICS_simple}
\end{multline}
with
\[
\mathcal Y_f(E_e)=\int_{E_e}^{m_{\chi}}
\left(\frac{{\rm d}N_e}{{\rm d}E_e'}\right)_f
\!{\rm Br}_f\,{\rm d}E_e'
\]
the number of $e^{\pm}$ produced per DM annihilation in channel $f$.

In our calculation we neglect diffusion, as done previously in Refs.~\cite{Cirelli:2009vg,DiMauro:2015tfa,Blanchet:2012vq,DiMauro:2023oqx}.
Indeed, the ratio of diffusion and energy-loss timescales is
\begin{equation}
\frac{\tau_{\rm diff}}{\tau_{\rm losses}}
=\frac{b(E_e)L^2}{E_e\,D(E_e)}
=\frac{b_0 L^2}{D_0}\,E_e^{\alpha-\delta-1},
\end{equation}
where $D(E_e)=D_0 E_e^{\delta}$ and $b(E_e)=b_0 E_e^{\alpha}$ are power-law approximations for diffusion and energy losses.
Numerically we find,
\begin{multline}
\frac{\tau_{\rm{diff}}}{\tau_{\rm{losses}}} = \frac{b_0 L^2}{D_0} E_e^{\alpha-\delta-1} \\
\approx \! 14 \! \left( \frac{b_0}{10^{-15} \, \rm{GeV/s}} \right) \!\left( \frac{10^{28} \, \rm{cm}^2/s}{D_0} \right) \times \\
\times \!\left(\frac{L}{4 \, \rm{kpc}}\right)^2 \!\left(\frac{E_e}{1 \,{\rm GeV}}\right)^{[0.2-0.6]}, \nonumber
\end{multline}
with $D_0=10^{28}\,\mathrm{cm^2\,s^{-1}}$, $\delta=0.4\text{--}0.6$, $b_0=10^{-15}\,\mathrm{GeV\,s^{-1}}$, and $\alpha=1.8\text{--}2.0$, values appropriate for the GC ISRF and magnetic fields at $E_e\gtrsim10\,\mathrm{GeV}$ \cite{Evoli:2016xgn}.
Thus $\tau_{\rm diff}\gg\tau_{\rm losses}$, \textit{i.e.}~energy losses are much more efficient than diffusion and the latter can be safely neglected.

As for the ISRF, we use the templates included in the {\tt GALPROP} cosmic-ray propagation code~\cite{Porter:2017vaa}, computed using two models.
The first, F98~\cite{1998ApJ...492..495F}, models emission from the Galactic bar and the stellar and dust disks, fitted to infrared surveys from \textsc{Cobe}/\textsc{Dirbe}.
The second, R12~\cite{2012A&A...545A..39R}, includes additional spatial components such as the bulge, halo, and spiral arms.
Overall, the two models give similar ISRF fluxes in our ROI, differing by factors of $\sim 0.1$–3 depending on photon energy.
We therefore adopt the F98 ISRF model for the rest of the paper.



\subsubsection{Bremsstrahlung photons}

Electrons and positrons produced in DM annihilations can also
emit bremsstrahlung photons via interactions with interstellar
gas.  Following Ref.~\cite{1999ApJ...513..311B} we compute this
contribution in analogy to Eq.~\ref{eq:ICS_simple}, replacing
the ICS power $\mathcal P_\text{ICS}$ by the bremsstrahlung kernel and
adopting a representative gas density
$n_{\rm gas}=1~\text{cm}^{-3}$ for the inner Galaxy
\cite{Evoli:2016xgn}.  For the channels and masses relevant to
the GCE, bremsstrahlung contributes below $\sim1$ GeV and is
subdominant to either the prompt signal ($b\bar b$) or the ICS
component ($\mu^+\mu^-,\tau^+\tau^-$).

Since diffusion (neglected in Eq.~\ref{eq:ICS_simple})
and bremsstrahlung affect the total flux in opposite directions
and largely cancel, we retain only
\emph{prompt + ICS} without diffusion as our baseline DM model.

\subsection{Upper limits using local dwarf galaxies}
\label{sec:dwarf}

dSphs are among the most promising targets for indirect DM searches due to their high mass-to-light ratios, low astrophysical backgrounds, and proximity to Earth~\cite{Abdo_2010,Ackermann:2015zua,Lopez:2015uma,Fermi-LAT:2016uux,%
Calore:2018sdx,Hoof:2018hyn,2019MNRAS.482.3480P,McDaniel:2023bju}. These satellite galaxies are believed to be DM-dominated systems, with dynamical mass estimates indicating DM densities several orders of magnitude larger than those in ordinary galaxies.

To estimate the upper limits from dSphs we use the list of objects, geometrical factors, and likelihood profiles reported in Ref.~\cite{McDaniel:2023bju}. If we consider only the prompt $\gamma$-ray emission, the resulting upper limits lie a factor of $\sim3$--5 above the best-fit values we obtain for the GCE. For example, in the $L_\mu-L_e$ model, fixing the DM mass to $43.8$~GeV and $m_{A'}=80$~GeV, we find an upper limit on $\langle \sigma v \rangle$ of $2\times10^{-25}\ \mathrm{cm}^3\,\mathrm{s}^{-1}$, while the best-fit to the GCE is about $(3.7$--$7.4)\times10^{-26}\ \mathrm{cm}^3\,\mathrm{s}^{-1}$ depending on the choice of GCE dataset and IEM systematics (see Fig.~\ref{fig:GCEresults}).

For models such as $L_i-L_j$ and $B-L$ that are dominated by leptonic channels, ICS is expected to be important. However, dSphs are located at distances of $\sim50$--200~kpc and have sizes of a few tens up to a few hundreds of parsecs. Energy losses are dominated by ICS on the cosmic microwave background, because the stellar and gas densities are too low to yield sizeable starlight or infrared radiation fields. As a consequence, the ratio of loss to diffusion timescales becomes
\[
\frac{\tau_{\rm loss}}{\tau_{\rm diff}} \approx 10^{-5}\text{--}10^{-4}\,
\left(\frac{10^{28}\ \mathrm{cm}^2\,\mathrm{s}^{-1}}{D_0}\right)
\left(\frac{E_e}{1~\mathrm{GeV}}\right)^{0.4}.
\]
Therefore, to include the ICS flux we need a precise estimate of $D_0$ around dSphs, since diffusion is the dominant process experienced by $e^{\pm}$ during their propagation in these systems. The strength of the diffusion coefficient in dSphs is highly uncertain and can vary between $10^{27}$ and $10^{29}\ \mathrm{cm}^2\,\mathrm{s}^{-1}$. In particular, for $D_0\gtrsim10^{27}\ \mathrm{cm}^2\,\mathrm{s}^{-1}$, electrons and positrons escape the dSph DM halo too quickly to generate a sizeable ICS flux, whereas $D_0\lesssim10^{27}\ \mathrm{cm}^2\,\mathrm{s}^{-1}$ is required to obtain strong ICS limits.

Ref.~\cite{Regis:2023rpm} studied turbulence self-generated by particles produced in DM annihilations. In their most optimistic scenario, they derive upper limits on the annihilation cross section for leptonic channels that are larger than our GCE best-fit values by a factor of a few. We therefore conclude that, even when adding the ICS flux to the prompt signal, the non-detection of $\gamma$-rays from dSphs does not constrain our model.

\subsection{{\sc Ams-02} cosmic positron flux}
\label{sec:ams02pos}

\begin{figure}
\centering
\includegraphics[width=0.99\linewidth]{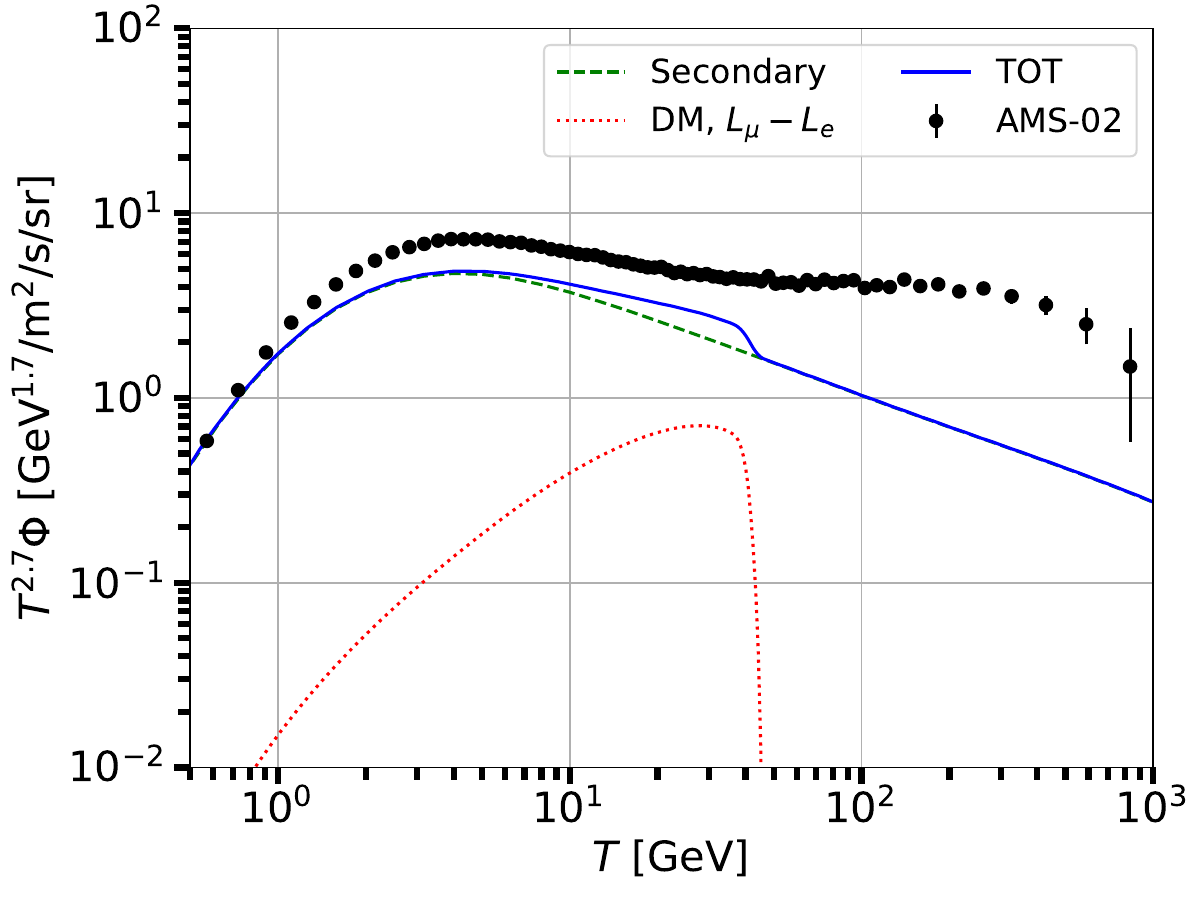}
    \caption{Flux of positrons from DM in the $L_e-L_\mu$ model for the best fit we obtain when fitting the GCE (dotter red curve). We also show the flux from the secondary production as reported in Ref.~\cite{DiMauro:2023oqx} (green dashed curve) and total positron flux (blue solid) compared to the {\sc Ams-02} data \cite{PhysRevLett.122.041102}.}
    \label{fig:posAMS}
\end{figure}

In the three models we consider, the dominant DM–annihilation channels are leptonic.  
Consequently, the ensuing electromagnetic shower produces copious positrons.  
We must therefore verify whether the DM distribution in the Milky Way halo could yield a cosmic‑ray positron flux detectable by {\sc Ams‑02}.

{\sc Ams‑02}, on board the International Space Station, has measured the positron spectrum with unprecedented precision over the energy range 0.5–1000 GeV~\cite{PhysRevLett.122.041102}.
At energies below 10 GeV the positron flux is dominated by \emph{secondary} particles created when primary cosmic rays undergo spallation on atoms in the interstellar medium~\cite{Orusa:2022pvp,DiMauro:2023oqx}.  
Theoretical uncertainties on this secondary component amount to $\sim10\%$ from the $e^{+}$ production cross‑sections~\cite{Orusa:2022pvp} and $\sim30\%$ from the poorly known vertical size of the diffusive halo.

Above 10 GeV, positrons produced in the relativistic winds of middle‑aged pulsars probably provide the dominant contribution to the flux (see, \textit{e.g.}, Refs.~\cite{DiMauro:2019yvh,DiMauro:2019hwn,Manconi:2020ipm,Orusa:2021tts,Orusa:2024ewq}).  
Predicting the cumulative pulsar contribution is, however, highly uncertain, because a handful of individual sources are expected to dominate the signal and each may have different electron–positron injection parameters that are difficult to constrain with multi‑messenger observations (see, \textit{e.g.}, Refs.~\cite{DiMauro:2019yvh,Manconi:2024wlq,VERITAS:2025xjd}).

To calculate the positron yield from DM annihilation we adopt same transport setup used in Ref.~\cite{DiMauro:2023oqx}, which is based on {\tt GALPROP}. We also consider the prediction of the secondary production from Ref.~\cite{DiMauro:2023oqx}, to which we refer all the details regarding the propagation model and $e^+$ secondary flux.
In order to evaluate conservative upper limits we fix the DM density distribution as the $\verb|MIN|$ configuration shown in Tab.~\ref{tab:GCEfit}.
An upper limit on $g_{X}$ is obtained by requiring that the total flux (secondary + DM) does not exceed any {\sc Ams‑02} data point by more than $2\sigma$.

We choose this approach rather than fitting the {\sc Ams‑02} data with a smooth analytic astrophysical template as in Ref.~\cite{Bergstrom:2013jra}, because the positron spectrum from DM in $L_{i}-L_{j}$ models closely resembles that expected from pulsars; employing an ad‑hoc functional form for the astrophysical background could therefore lead to artificially stringent constraints.

When fixing for each model the value of the DM mass obtained from the fit to the GCE (see Tab.~\ref{tab:GCEfit}) we obtain an upper limit for $\langle \sigma v \rangle$ that is $1.75 \times 10^{-25}$ cm$^3$/s, $5.4 \times 10^{-26}$ cm$^3$/s, $1.12 \times 10^{-25}$ cm$^3$/s and $1.20 \times 10^{-25}$ cm$^3$/s for the $L_e-L_\mu$, $L_e-L_\tau$, $L_\mu-L_\tau$ and $B-L$, respectively. These values are well above the values needed to fit the GCE.
We show in Fig.~\ref{fig:posAMS} the comparison of the total flux coming from DM, when choosing the $L_e-L_\mu$ model for the best fit values of DM mass and $\langle \sigma v\rangle$ we obtain by fitting the GCE, and secondary production with the {\sc Ams-02} data. The predicted flux is well below the data.

\section{Results}
\label{sec:results}

\subsection{Fit to the Galactic Center Excess}
\label{sec:resultsGCE}

\begin{figure*}
\centering
\includegraphics[width=0.49\linewidth]{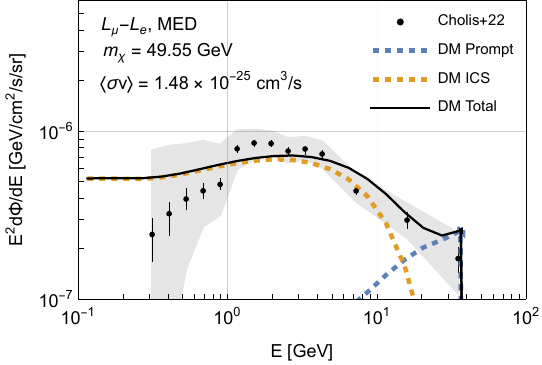}
\includegraphics[width=0.49\linewidth]{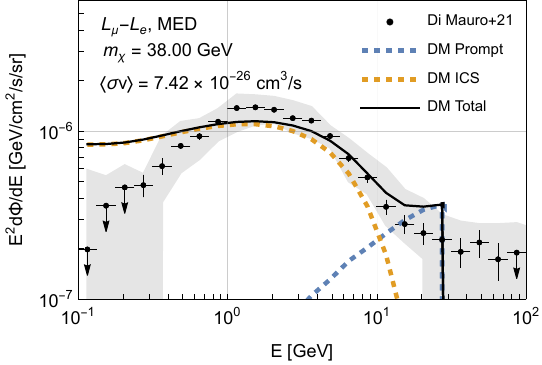}
\includegraphics[width=0.49\linewidth]{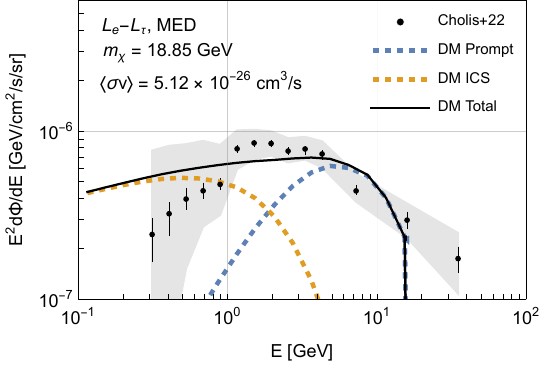}
\includegraphics[width=0.49\linewidth]{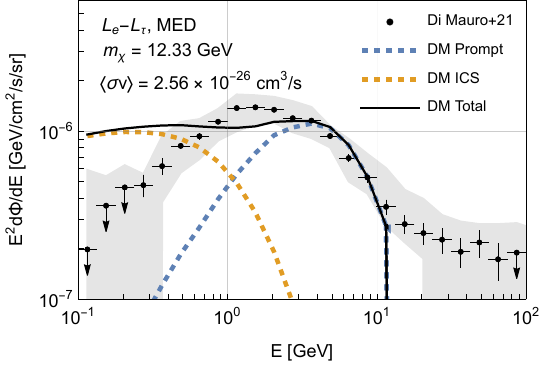}
\includegraphics[width=0.49\linewidth]{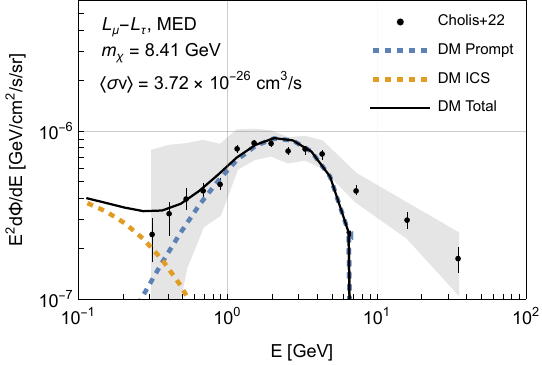}
\includegraphics[width=0.49\linewidth]{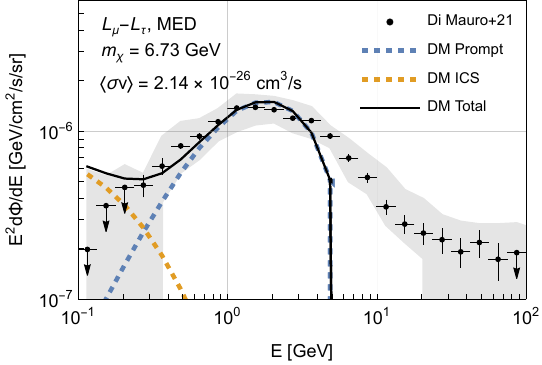}
\includegraphics[width=0.49\linewidth]{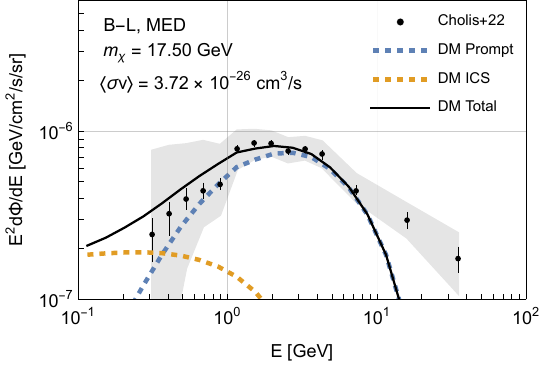}
\includegraphics[width=0.49\linewidth]{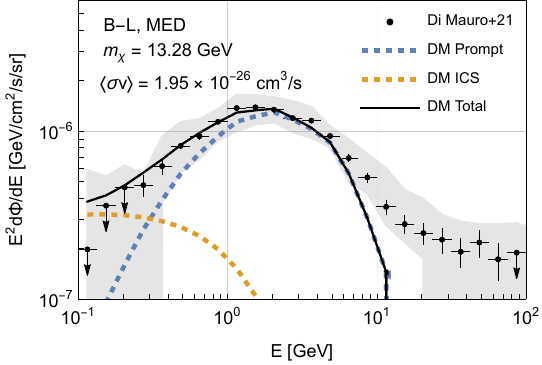}
    \caption{Best-fit of the four models considered in this work to the Di Mauro+21 and Cholis+22 GCE datasets. In each figure we report the GCE data and the theoretical predictions for the prompt (blue curve), ICS (red curve) and total $\gamma$-ray flux (black curve).}
    \label{fig:GCEresults}
\end{figure*}

We fit the GCE data reported by Di Mauro+21 and Cholis+22 \cite{DiMauro:2021qcf,Cholis:2021rpp}. As data uncertainties we adopt the envelope spanned by the GCE spectra obtained when varying the IEM templates; indeed, these systematics dominate over statistical errors. The fit results for each model and data set are listed in Tab.~\ref{tab:GCEfit} and illustrated in Fig.~\ref{fig:GCEresults}.

For all models, the annihilation cross section required to fit the GCE is remarkably close to the thermal value that yields the correct relic abundance. The DM mass lies in the standard WIMP range, $m_\chi \simeq 7$–$50\,\mathrm{GeV}$. Moreover, the best-fit values of $m_\chi$ and $\langle \sigma v \rangle$ obtained from the two data sets are broadly compatible. In particular, the $\langle \sigma v \rangle$ preferred by Cholis+22 is larger by about a factor of two compared to Di Mauro+21. There are two main reasons: (i) the best-fit $m_\chi$ is systematically higher for Cholis+22, which drives $\langle \sigma v \rangle$ upward, and (ii) the different ROI choices—most notably the masking of the Galactic disk by Cholis+22.—reduce the inferred GCE normalization.

The two models that provides the best fit are $L_{\mu}-L_e$ and $B-L$. In the former BSM scenario the ICS component, which better reproduces the GCE peak at a few GeV, dominates over the prompt emission. For the Di Mauro+21 fit, the predicted flux exceeds the uncertainty band below $300\,\mathrm{MeV}$; however, that energy range is strongly affected by IEM systematics.
In the $B-L$ model DM annihilates significantly into quark pairs. As it is visible in Fig.~\ref{fig:GCEresults}, BSM with DM hadronic annihilation channels have a prompt flux that is much larger than the ICS one and that can fit properly the GCE spectrum.
In the $L_{e}-L_\tau$ model, the prompt and ICS components have comparable intensities in $E^2\,{\rm d}\Phi/{\rm d}E$, but their spectra peak at energies differing by roughly an order of magnitude. Their sum therefore provides a poorer fit than in the previous case because the data have a single peak. Finally, in the $L_{\mu}-L_\tau$ model the prompt emission dominates over ICS; its narrower peak matches the low-energy data and the main bump but fails to reproduce the high-energy tail above $\sim 5\,\mathrm{GeV}$, where the flux is cut off kinematically by $m_\chi$.

We also note the slight impact on the fit to the GCE when using the R12 ISRF model instead of the F98 one. 
For example, for the $L_\mu - L_e$ model we obtain
$m_\chi = 41.35\,(50.07)\,\mathrm{GeV}$ and
$\langle\sigma v\rangle = 3.52\times10^{-26}\,(6.16\times10^{-26})\;\mathrm{cm}^{3}\,\mathrm{s}^{-1}$,
with $\chi^{2}/\text{d.o.f.}=1.73\,(0.04)$ when using the Di Mauro+21 (Cholis+22) dataset. 
Considering that the systematic uncertainties associated with the choice of the IEM and the DM density profile affect the GCE fit more strongly than the choice of the ISRF model, we therefore neglect the impact of the latter, as stated in Sec.~\ref{sec:observables}.

\begin{table}[]
    \centering
    \begin{tabular}{|c|c|c|c|c|}
    \hline
     Model & Dataset & $m_\chi$ [GeV] & $\langle\sigma v\rangle$ [cm$^3$/s] & $\chi^2$/d.o.f. \\
    \hline
    \multirow{2}{*}{$L_\mu-L_\tau$} & Di Mauro+21 & $6.73$ & $2.14\times 10^{-26}$ & $3.84$ \\
    \cline{2-5}
    & Cholis+22 & $8.41$ & $3.72\times 10^{-26}$ & $0.41$ \\
    \hline
    \multirow{2}{*}{$L_e-L_\tau$} & Di Mauro+21 & $12.33$ & $2.56\times 10^{-26}$ & $4.68$ \\
    \cline{2-5}
    & Cholis+22 & $18.85$ & $5.12\times 10^{-26}$ & $0.28$ \\
    \hline    
    \multirow{2}{*}{$L_\mu-L_e$} & Di Mauro+21 & $38.00$ & $7.42\times 10^{-26}$ & $2.12$ \\
    \cline{2-5}
    & Cholis+22 & $49.55$ & $1.48\times 10^{-25}$ & $0.05$ \\
    \hline
    \multirow{2}{*}{$B-L$} & Di Mauro+21 & $13.28$ & $1.95\times 10^{-26}$ & $1.99$ \\
    \cline{2-5}
    & Cholis+22 & $17.50$ & $3.72\times 10^{-26}$ & $0.31$ \\  
    \hline
    \end{tabular}
    \caption{Results of the fits to the GCE of the DM mass $m_\chi$ and thermally averaged annihilation cross section $\langle\sigma v\rangle$, for the three $L_i-L_j$ models and the $B-L$ model and the GCE datasets reported in Refs.~\cite{DiMauro:2021qcf,Cholis:2021rpp}}
    \label{tab:GCEfit}
\end{table}

\subsection{Combined results}

\begin{figure*}[]
\centering
\includegraphics[width=0.49\linewidth]{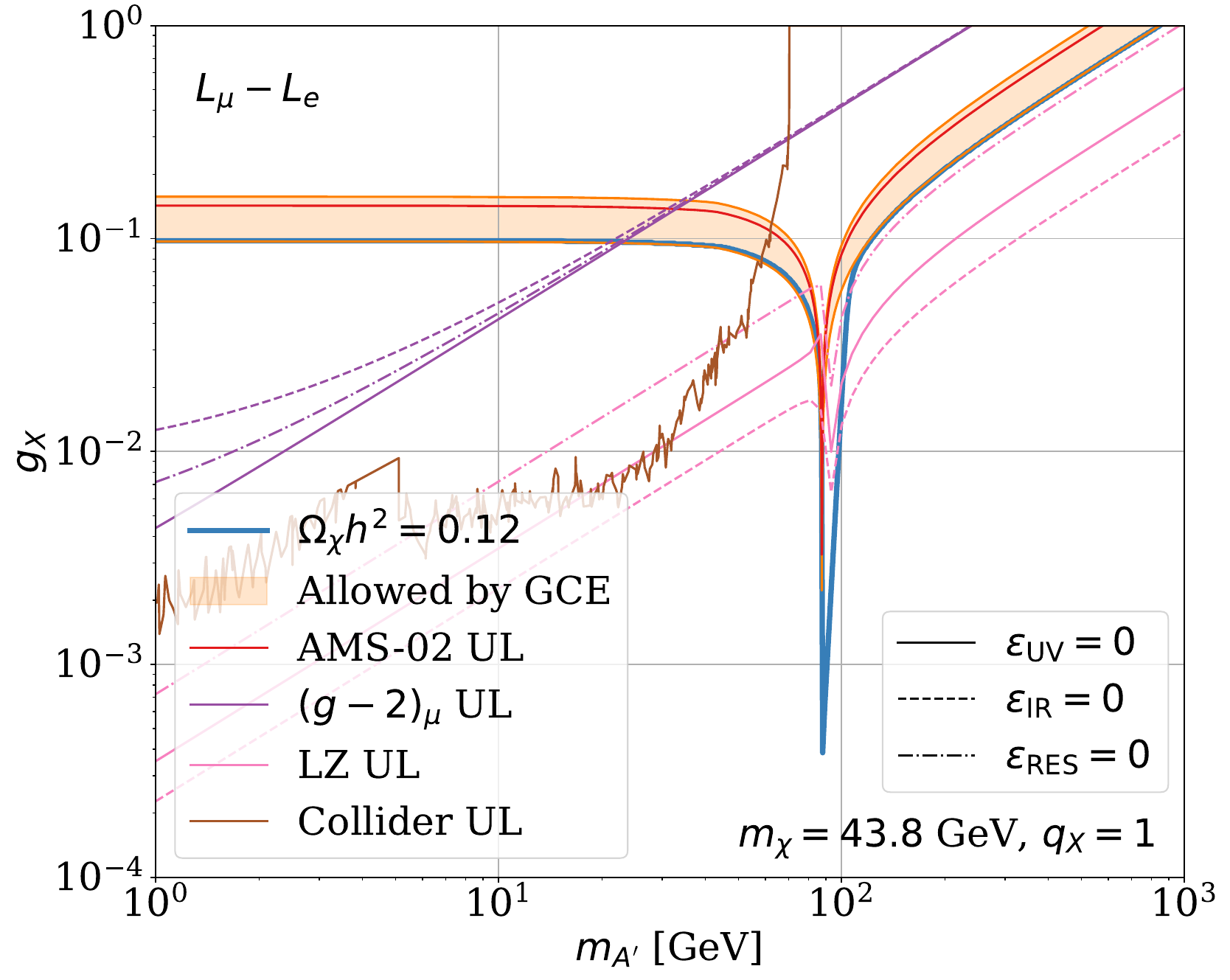}
\includegraphics[width=0.49\linewidth]{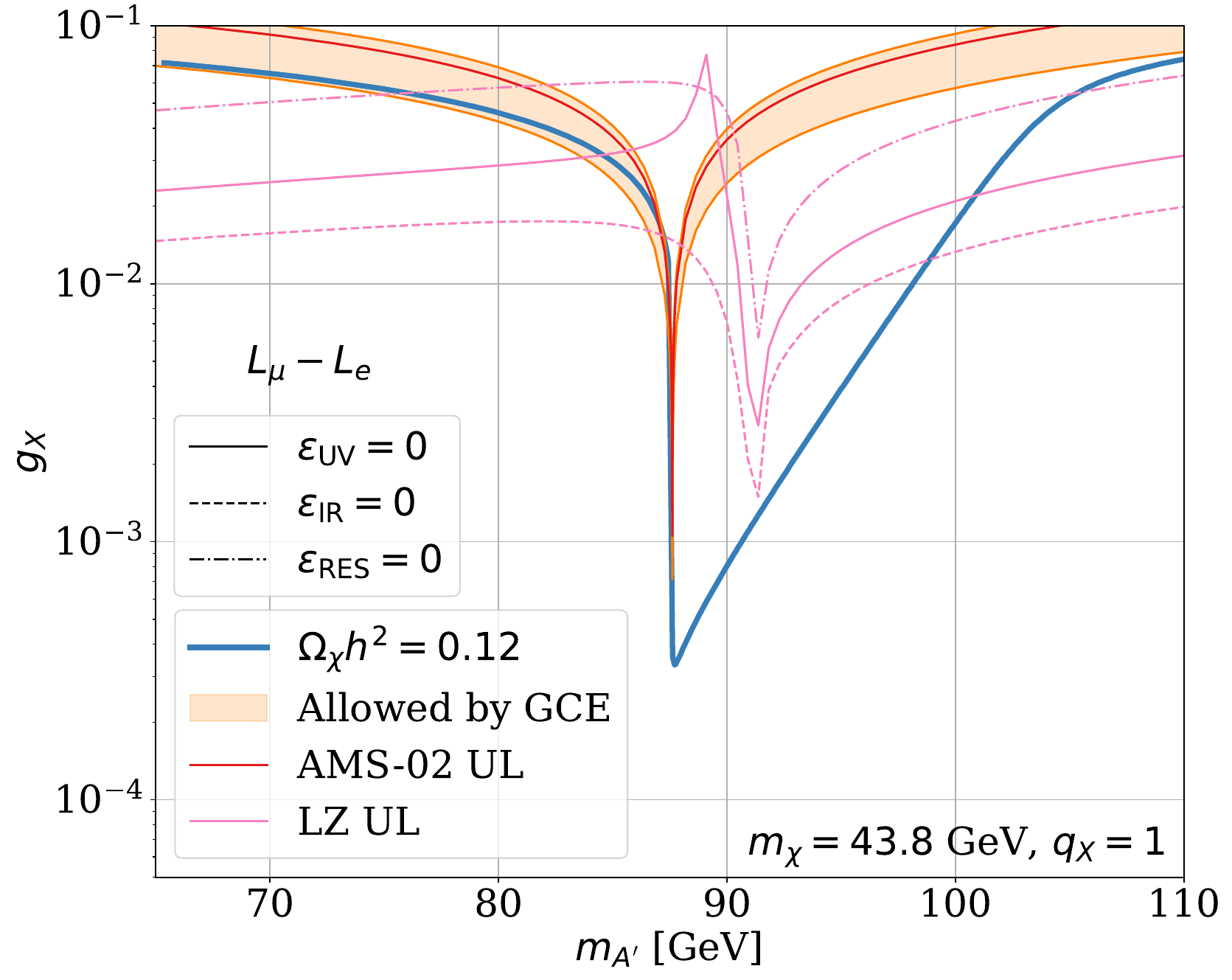}
\includegraphics[width=0.49\linewidth]{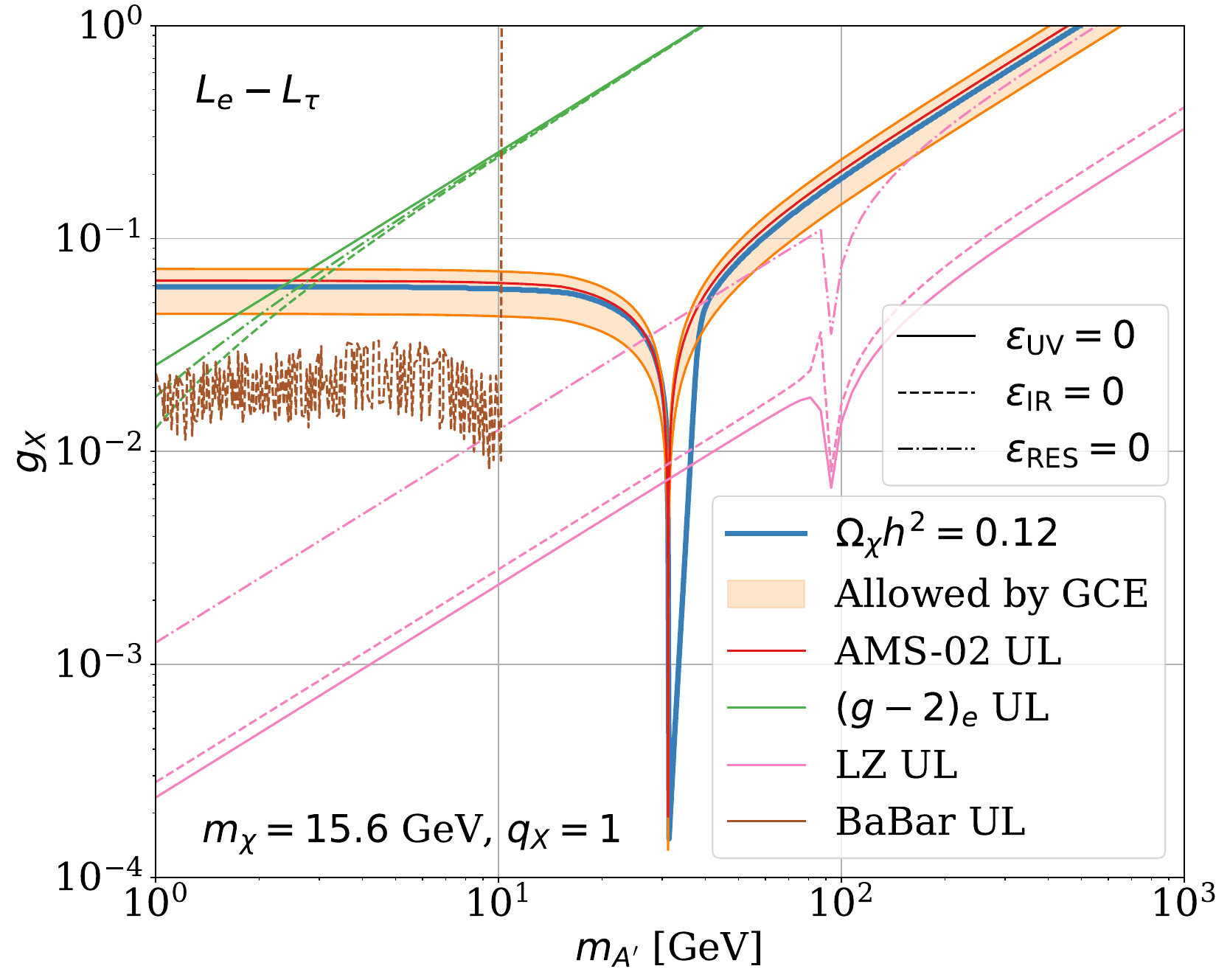}
\includegraphics[width=0.49\linewidth]{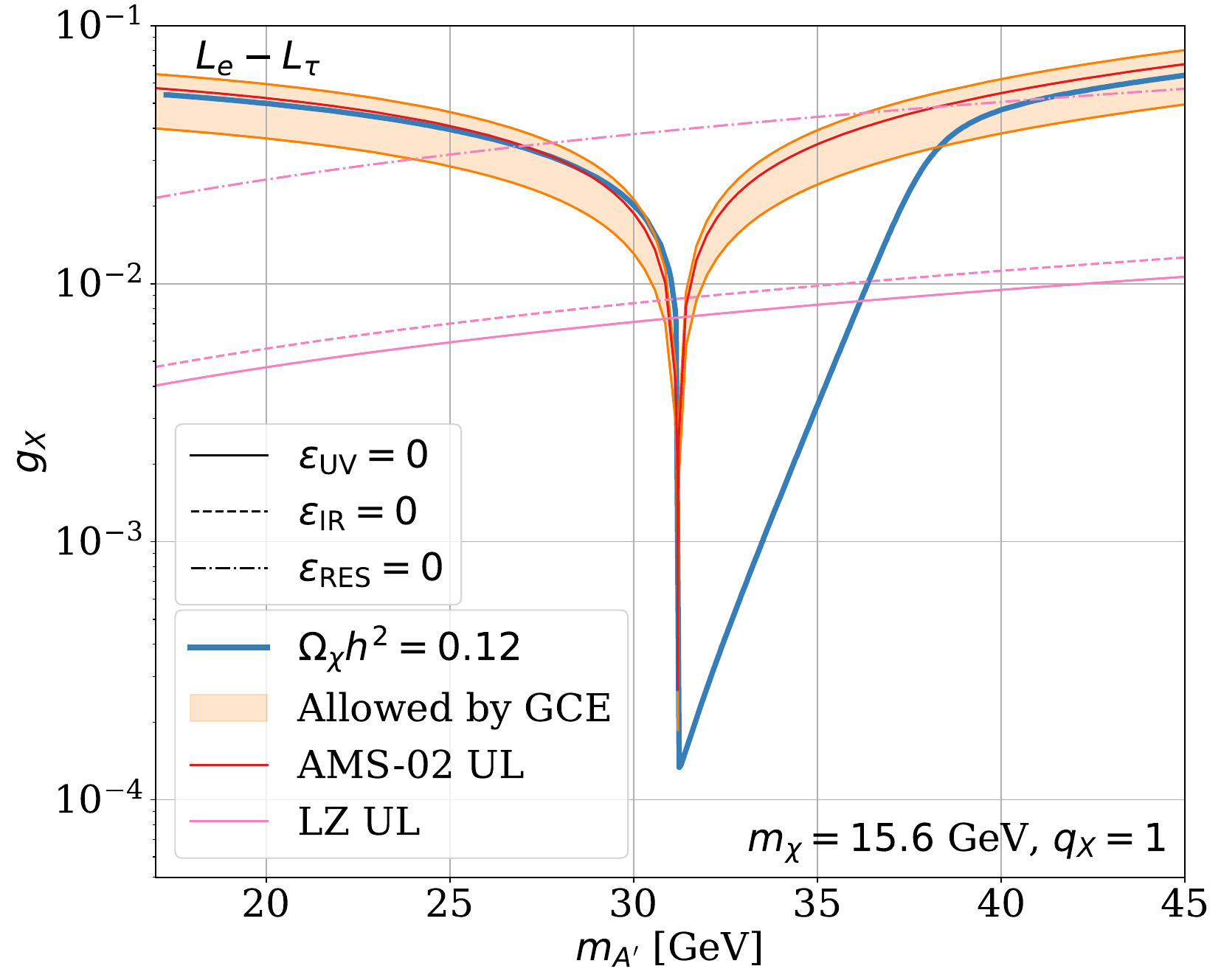}
\includegraphics[width=0.49\linewidth]{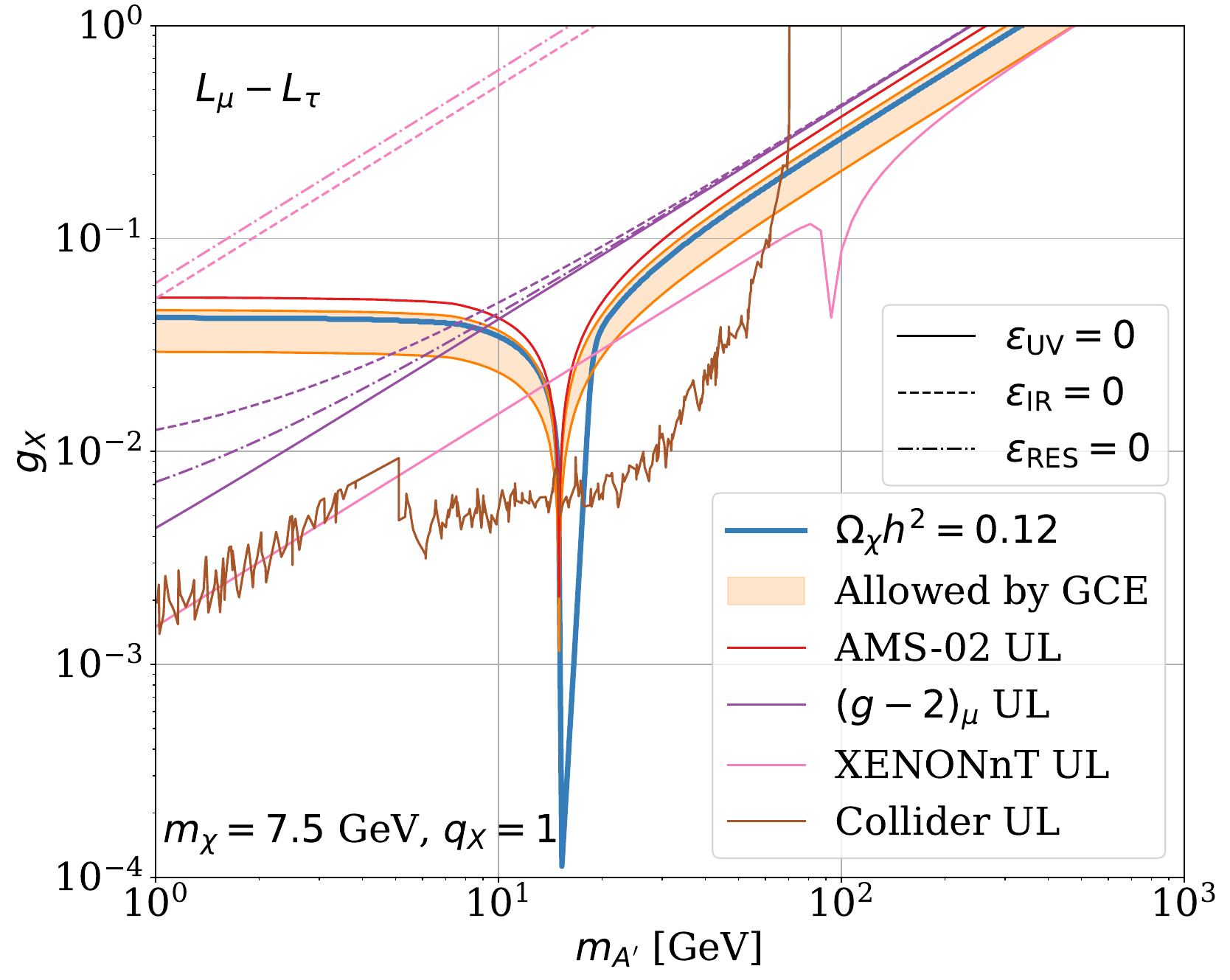}
\includegraphics[width=0.49\linewidth]{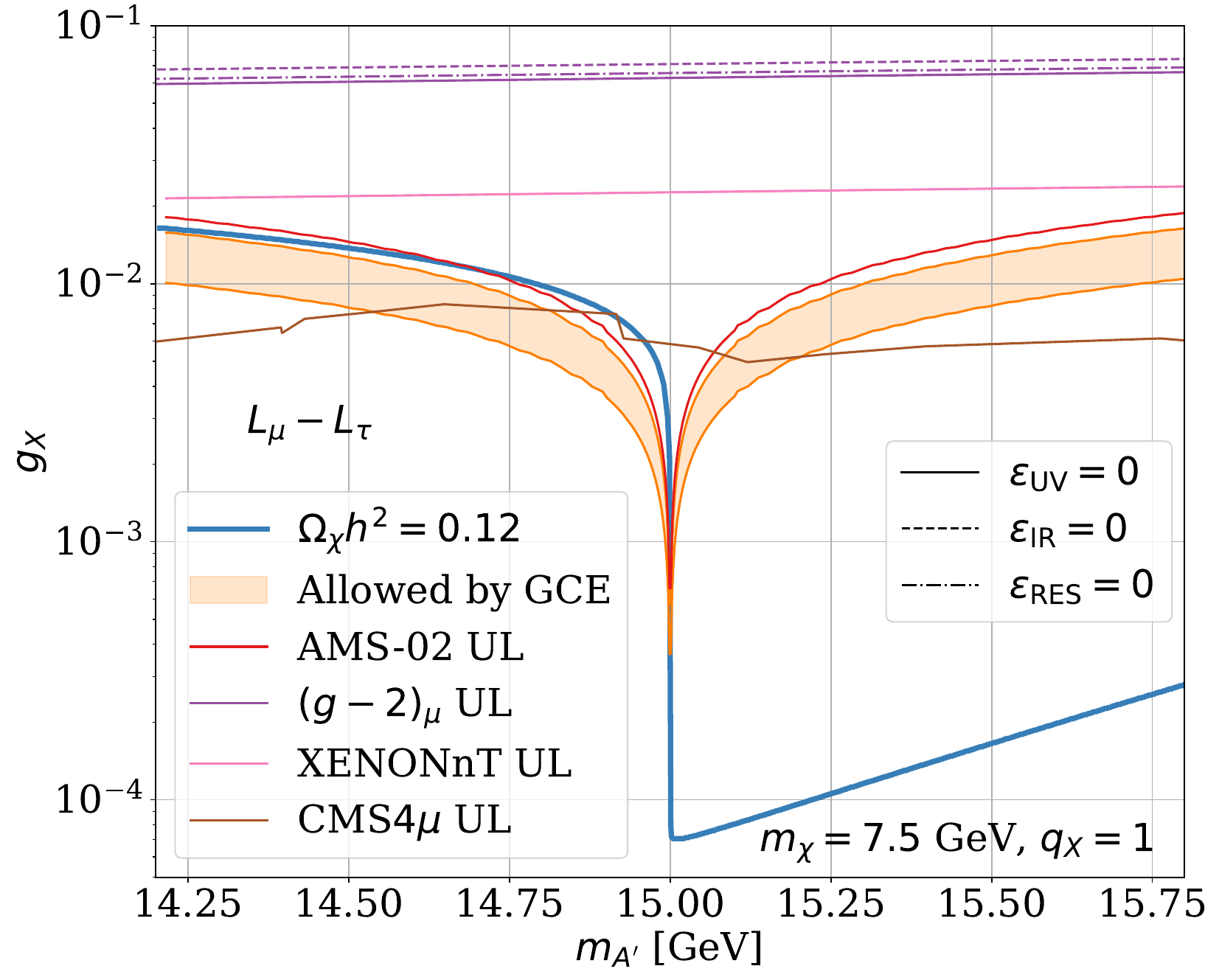}
    \caption{For each of the $U(1)_{L_i-L_j}$, constraints on the $g_X/m_{A'}$ parameter space for $m_\chi$ providing the best-fit to the GCE. The top panels show the constraints for $L_\mu-L_e$, the middle ones for $L_e-L_\tau$ and the bottom ones for $L_\mu-L_\tau$. The left panels show a broader range of $m_{A'}$, while the right panels are showing the resonance region (when $m_{A'} \sim 2m_\chi$). We show: the parameter space for which the thermal relic abundance reaches $\Omega_\chi h^2=0.12$ (solid blue) and for which the GCE is explained by DM (orange). Furthermore, we show different upper limits (UL): from the positron flux measured by {\sc Ams-02} (solid red), from colliders, especially from {\sc BaBar}, {\sc Cms} and {\sc Ccfr} (brown), from $(g-2)_e$ (green) and $(g-2)_\mu$ (purple) and finally from DD, especially from {\sc Lz} and {\sc Xenon}n{\sc T} (pink). We also showcase the limits for different kinetic mixing settings: $\epsilon_\text{UV}=0$ (solid), $\epsilon_\text{IR}=0$ (dashed) and $\epsilon_\text{RES}=0$ (dot-dashed).}
    \label{fig:results}
\end{figure*}

\begin{figure*}[]
\centering
\includegraphics[width=0.49\linewidth]{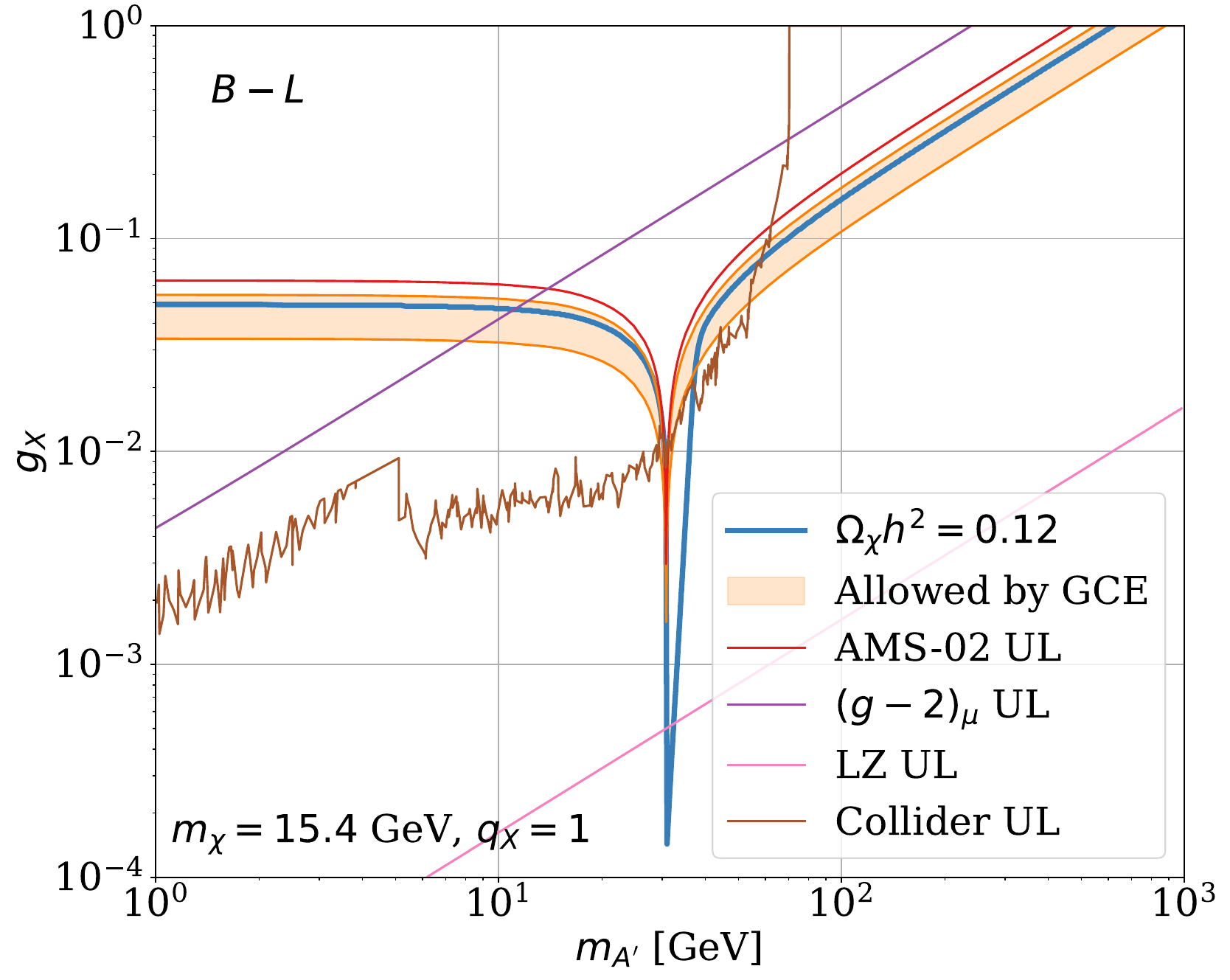}
\includegraphics[width=0.49\linewidth]{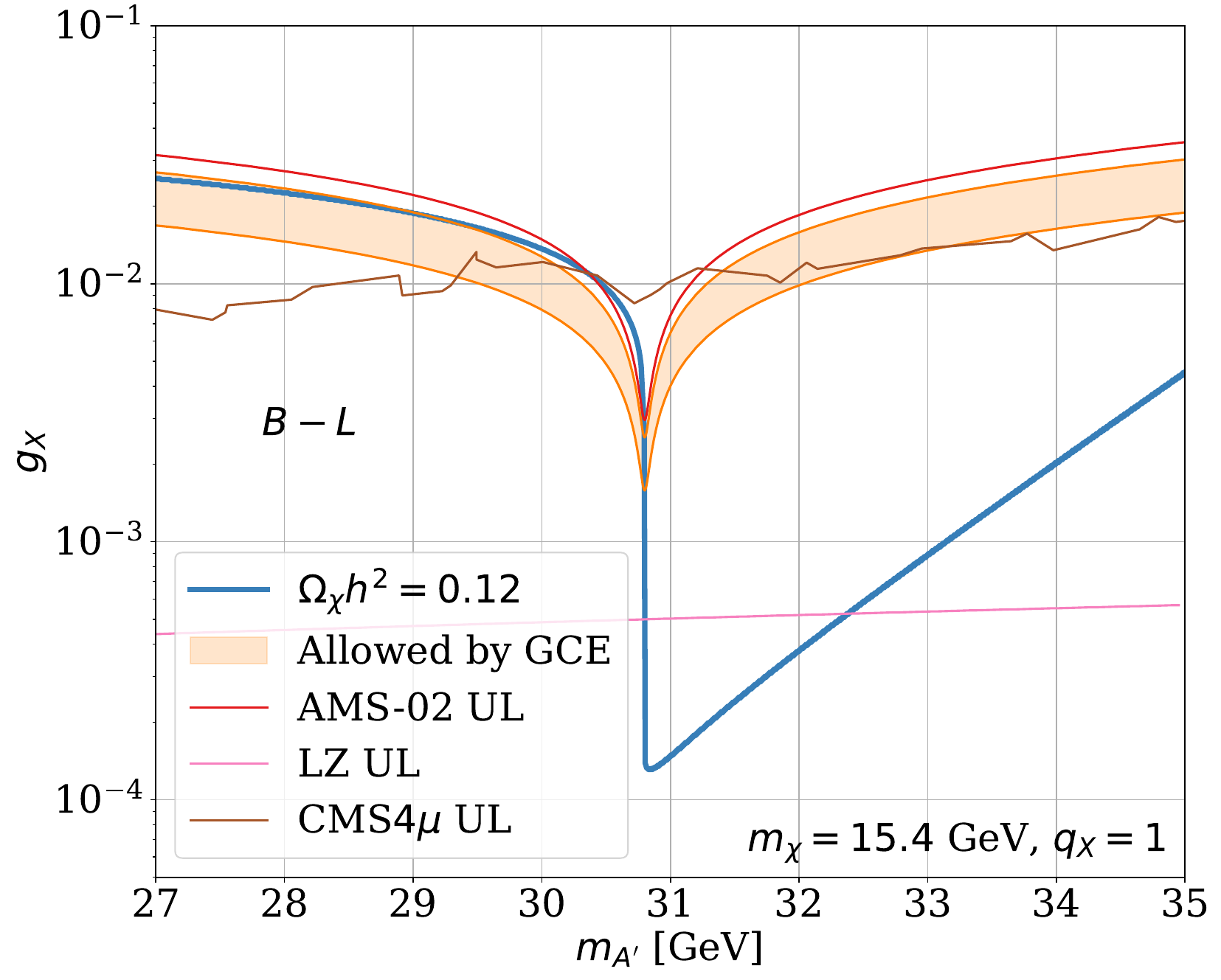}
    \caption{Same as Fig.~\ref{fig:results}, but for the $U(1)_{B-L}$ model.}
    \label{fig:resultsBL}
\end{figure*}

In this section we report the combined results we obtain for the models $U(1)_{L_i-L_j}$ in Fig.~\ref{fig:results} and for $U(1)_{B-L}$ in Fig.~\ref{fig:resultsBL}.

For each model we display the region that fits the GCE spectrum, shown as a band that incorporates the uncertainties in the best‑fit values of $\langle\sigma v\rangle$ and the DM mass derived from the Di~Mauro+21 and Cholis+22 data sets.  
We also include the variation induced by the choice of the DM‑density parameters that enter the geometrical factor.  
The uncertainty on $\overline{\mathcal{J}}$ is dominant: it translates into roughly a factor 2 in the predicted $\gamma$‑ray flux and about a 20\% uncertainty on $g_X$ (see Tab.~\ref{tab:Jfact}).

We further show the portion of parameter space that reproduces the observed relic density.  
As expected, the values of $g_X$ that satisfy both the GCE and $\Omega_\chi h^{2}$ are nearly constant for $m_{A'} < 2m_\chi$, lying between 0.01 and 0.1 depending on the model.  
Above the resonance, $g_X$ grows approximately linearly with $m_{A'}$, whereas in the vicinity of the resonance ($m_{A'} \simeq 2m_\chi$) it dips sharply because the annihilation cross‑section rises and the coupling must decrease to keep $\langle\sigma v\rangle \simeq (1-3)\times10^{-26}\,\mathrm{cm^{3}\,s^{-1}}$.

Constraints from the anomalous magnetic moments of the muon and the electron, $(g-2)_\mu$ and $(g-2)_e$, are overlaid for the three benchmark choices $\epsilon_\mathrm{UV}=0$, $\epsilon_\mathrm{IR}=0$, and $\epsilon_\mathrm{RES}=0$ for $L_i-L_j$ models.  
Tree‑level kinetic mixing is relevant only for $m_{A'} \lesssim 10\,\mathrm{GeV}$.  
For $\epsilon_\mathrm{UV}=0$ the limit is a straight line in log–log space at low $m_{A'}$ because the tree‑level kinetic mixing in Eq.~\ref{eq:dala} vanishes.  
In the other two cases $\epsilon$ is non‑zero (larger for $\epsilon_\mathrm{IR}=0$ than for $\epsilon_\mathrm{RES}=0$), so the bounds are correspondingly weaker.

Collider constraints matter chiefly below $m_{A'} \lesssim 10\,\mathrm{GeV}$ for the $L_e-L_\tau$ model, where we include limits from {\sc BaBar}~\cite{BaBar:2014zli} (converted from the limits on $\epsilon$ for the $\epsilon_\text{IR} = 0$ case).
For the $B-L$, $L_\mu-L_\tau$ and $L_e-L_\mu$ models we show the (muon-only) {\sc BaBar}~\cite{BaBar:2016sci} and {\sc Ccfr}~\cite{CCFR:1991lpl} limits for $m_{A'} \lesssim 10\,\mathrm{GeV}$ as well as the {\sc Cms} limit, which extends up to $m_{A'} \simeq 80\,\mathrm{GeV}$~\cite{CMS:2018yxg}.

Finally, we overlay the nuclear‑recoil limits from {\sc Xenon}n{\sc T} and {\sc Lz} for the same three $\epsilon$ benchmarks.  
As expected, taking $\epsilon_\mathrm{RES}$ weakens the constraint on $g_X$ by about a factor of two for $L_e-L_\mu$, and even more for the other two $L_i-L_j$ models.

The $B-L$ model is a typical BSM scenario in the WIMP framework.
Here DM particles annihilate into pairs of SM fermions, and the observed relic density can be reproduced for couplings 
$g_X$ of the same order as the electroweak ones, $g_X\!\sim\!(g,g')$.
Moreover, direct detection proceeds at tree-level.
Current \textsc{Xenon}n{\sc T} and \textsc{Lz} nuclear-recoil limits on $\sigma_N$ already exclude most of the model’s parameter space.
The only region that survives is a narrow, fine-tuned strip near resonance, where the DM mass is within a few \% of half the mediator mass.
Next-generation experiments such as \textsc{Darwin}, expected to reach the neutrino floor, will almost completely probe this remaining parameter space~\cite{DARWIN:2016hyl}.

Instead, $L_i-L_j$ models do not suffer as the $B-L$ for such tight limits from direct detection. In fact, the upper limits on $g_X$ from $\sigma_N$ are about factor of 100 weaker.
This implies that a broader region around the resonance can satisfy the relic density and direct detection constraints.

$L_e-L_\mu$ model provides the best description of the GCE with a DM mass around $40$–$50\,\mathrm{GeV}$, depending on the data set, and a cross‑section close to the thermal value.  
Considering the least constraining $\epsilon_{\rm{RES}}$ case, direct detection data exclude this model for $m_{A'} \lesssim 70\,\mathrm{GeV}$ and $m_{A'} \gtrsim 105\,\mathrm{GeV}$.  
Within the intermediate range there is a viable window $m_{A'}\in[70,87]\,\mathrm{GeV}$ in which all current bounds (GCE, relic density, collider, {\sc Ams‑02} $e^+$ flux, and direct detection) are satisfied.  
A tiny corner near $m_{A'}\simeq105\,\mathrm{GeV}$ and $g_X\simeq 5\times10^{-2}$ also fits the GCE and relic density but is only marginally excluded by direct detection.

The $L_\mu-L_\tau$ and $L_e-L_\tau$ models yield a poorer fit to the GCE.  Moreover, except very close to the resonance, no region simultaneously matches the GCE, relic density, and all current constraints.  
An exception occurs for $L_e-L_\tau$ at $m_{A'}\simeq37$–$38\,\mathrm{GeV}$ and $g_X\simeq(2$–$4)\times10^{-2}$, which remains compatible with all bounds while fitting both the GCE and the relic density.

\section{Conclusions}
\label{sec:conclusions}

We have performed a global analysis of DM coupled to a gauge boson in the four anomaly‑free realisations \(U(1)_{L_\mu-L_e}\), \(U(1)_{L_\mu-L_\tau}\), \(U(1)_{L_e-L_\tau}\) and \(U(1)_{B-L}\) confronting each scenario with the most recent data from the GCE, relic‑density calculations, direct detection experiments, collider searches, anomalous magnetic moments, and the {\sc Ams‑02} positron spectrum. 
The first three models, labeled as $L_i-L_j$, contain a mediator that is leptophilic, \textit{i.e.}~it decays mostly into lepton pairs, and the direct detection cross section is suppressed by the kinetic mixing term. Instead in the $B-L$ model, DM annihilates significantly also into quarks and direct detection occurs at tree-level without kinetic mixing suppression.
Once prompt emission and ICS are included, the GCE can be reproduced in all four cases with a velocity‑averaged annihilation cross section very close to the canonical thermal value \(3\times10^{-26}\,\text{cm}^{3}\,\text{s}^{-1}\).  
Fitting the Di Mauro+21 and Cholis+22 datasets yields best‑fit DM masses of \(13\)–\(18\;\text{GeV}\) for \(B-L\), \(6\)–\(9\;\text{GeV}\) for \(L_\mu-L_\tau\), \(12\)–\(19\;\text{GeV}\) for \(L_e-L_\tau\) and \(38\)–\(50\;\text{GeV}\) for \(L_\mu-L_e\). The latter model is the one that provides the best match with the GCE data. 

Compatibility with the relic density and with laboratory constraints forces the mediator mass to sit close to the \(s\)-channel resonance, \(m_{A'}\simeq2m_\chi\), particularly for the $B-L$, $L_\mu-L_\tau$ and $L_e-L_\tau$.
In that narrow region the value of the coupling $g_X$ is around \(g_{X}\simeq(1\text{--}4)\times10^{-2}\) and falling below \(10^{-2}\) only when the resonance condition is satisfied at the per‑cent level, as for the $B-L$ model.  
For \(U(1)_{L_\mu-L_e}\) a particularly robust window survives, \(m_{A'}\simeq70\text{--}86\;\text{GeV}\) with \(g_{X}\simeq0.01\text{--}0.05\); a tiny corner near \(m_{A'}\simeq105\;\text{GeV}\) and \(g_{X}\simeq5\times10^{-2}\) also remains viable.
The \(U(1)_{L_e-L_\tau}\) model retains a narrow strip around \(m_{A'}\simeq37\text{--}38\;\text{GeV}\) and \(g_{X}\simeq(2\text{--}4)\times10^{-2}\), while \(U(1)_{L_\mu-L_\tau}\) is allowed only for \(m_{A'}\simeq14\text{--}16\;\text{GeV}\) (\textit{i.e.}~very close to the resonance) and for couplings of the order of $10^{-4}-10^{-3}$.
The $B-L$ model is excluded in most of the parameter space except when the DM mass is equal to half the mediator mass with a fine tune of a few \%.
Outside these resonant bands direct detection limits from {\sc Xenon}n{\sc T} and {\sc Lz} are dominant below the resonance, collider searches exclude large couplings above \(\sim10\;\text{GeV}\), and \((g-2)_{e,\mu}\) constrains the sub‑GeV region.

Among the three gauge choices, \(U(1)_{L_\mu-L_e}\) gives the best overall description of the data: it fits both GCE spectra, satisfies the relic‑density requirement, and evades all current bounds once the kinetic mixing suppression is taken into account but without a finetune of the DM and mediator mass.  
A factor‑ten improvement in Xenon‑based nuclear‑recoil sensitivity will prove a factor of 2 lower values for $g_X$ but it cannot exclude the region very close to the resonance.


\medskip

\begin{acknowledgments} 
M.D.M.~and J.K.~are grateful to Giorgio Arcadi and Nicolao Fornengo for their helpful feedback on the draft. They thank Patrick Foldenauer for providing the \verb|Feynrules| model files. They also acknowledge support from the research grant {\sl TAsP (Theoretical Astroparticle Physics)} funded by Istituto Nazionale di Fisica Nucleare (INFN) and from the Italian Ministry of University and Research (MUR), PRIN 2022 ``EXSKALIBUR – Euclid-Cross-SKA: Likelihood Inference Building for Universe’s Research'', Grant No. 20222BBYB9, CUP I53D23000610 0006, and from the European Union -- Next Generation EU.
\end{acknowledgments}

\bibliographystyle{apsrev4-1}
\bibliography{paper.bib}

\newpage

\onecolumngrid
\appendix

\section{Deriving the mass eingestates of $L_i-L_j$ models}
\label{app:massstates}

In this appendix, we derive the interaction Lagrangian for a gauged $L_i - L_j$ model ($L_e - L_\mu$, $L_e - L_\tau$ or $L_\mu - L_\tau$) in detail. 
In particular, we explain the procedure how, from the full gauge–kinetic Lagrangian with kinetic mixing we obtain the mass eigenstates. We first apply a rotation to remove the kinetic mixing term and obtain the canonical forms of the field strength tensors for $B$ and the new boson $X$. Then, we apply the rotation to the mass eigenstate basis in order to finally derive the interaction Lagrangian between each physical gauge boson to SM and dark sector currents.

We begin with the gauge-kinetic Lagrangian involving the Standard Model (SM) $U(1)_Y$ (hypercharge) field $\hat B_\mu$, the $SU(2)_L$ neutral gauge field $\hat W^3_\mu$, and the new $U(1)_{L_i-L_j}$ gauge field $\hat X_\mu$. The Lagrangian of the model, that includes a kinetic mixing term and mass terms, is written as: 
\begin{equation}\label{eq:Lag_start}
\mathcal{L} \;\supset\; -\frac{1}{4}\,\hat B_{\mu\nu}\hat B^{\mu\nu} \;-\;\frac{1}{4}\,\hat W^3_{\mu\nu}\hat W^{3\,\mu\nu} \;-\;\frac{1}{4}\,\hat X_{\mu\nu}\hat X^{\mu\nu} \;-\; \frac{\epsilon}{2}\,\hat B_{\mu\nu}\hat X^{\mu\nu} \;-\; g'J_Y^\mu \hat B_\mu \;-\; g\,J_3^\mu \hat W^3_\mu \;-\; g_X\,J_X^\mu \hat X_\mu \;+\; \frac{1}{2}m_X^2\,\hat X_\mu \hat X^\mu~,
\end{equation}
where $\hat B_{\mu\nu}, \hat W^3_{\mu\nu}, \hat X_{\mu\nu}$ are the field strength tensors for $\hat B_\mu, \hat W^3_\mu, \hat X_\mu$ respectively, and $\epsilon$ is the kinetic mixing parameter (assumed small, $\epsilon \ll 1$). The gauge couplings are denoted by $g'$ (for hypercharge $U(1)_Y$), $g$ (for $SU(2)_L$), and $g_X$ (for the new $U(1)_{L_i-L_j}$). The currents $J_Y^\mu$, $J_3^\mu$, and $J_X^\mu$ are the matter currents associated with hypercharge, $SU(2)_L$ third-component, and $L_i-L_j$ charges, respectively. In particular, $J_Y^\mu = \sum_f Y_f\,\bar f \gamma^\mu f$ (summing over all fermions $f$ with hypercharge $Y_f$), $J_3^\mu = \sum_f T^3_f\,\bar f \gamma^\mu f$ (with $T^3_f$ the third-component of isospin), and $J_X^\mu = \sum_f q_{X,f}^f\,\bar f \gamma^\mu f + q_{\chi} \bar{\chi} \gamma^\mu \chi$ for the new charge $q_X$. In a $L_i - L_j$ model, the charge assignments are such that $q_{X,i}=+1$ for all leptons of flavor $i$ (and their corresponding neutrinos), $q_{X,j}=-1$ for all leptons of flavor $j$, and $q_{X,k}=0$ for every other SM field. 
For example, in a $L_\mu - L_\tau$ model, $\mu^-,\nu_\mu$ carry $q_{X,\mu}=+1$, $\tau^-,\nu_\tau$ carry $q_{X,\tau}=-1$, and all other fermions have $q_{X,k}=0$. The current $J_X^\mu$ also includes contributions from a new fermions $\chi$ carrying a $L_i-L_j$ charge $q_X(\chi)=1$.

The first step is to eliminate the kinetic mixing term $-\frac{\epsilon}{2}\hat B_{\mu\nu}\hat X^{\mu\nu}$ and achieve canonical normalization of all gauge kinetic terms. This can be done by a linear field redefinition (which is non-orthogonal since it involves mixing of two $U(1)$ fields). Specifically, we introduce new basis fields $B_\mu$ and $X_\mu$ defined by: 
\begin{equation}\label{eq:kin_mix_transform}
B_\mu \;=\; \hat B_\mu + \epsilon\,\hat X_\mu~, \qquad\qquad 
X_\mu \;=\; \sqrt{\,1-\epsilon^2\,}\;\hat X_\mu~,
\end{equation}
while leaving $\hat W^3_\mu$ unchanged ($W^3_\mu \equiv \hat W^3_\mu$). To all orders in $\epsilon$, one can verify that this transformation removes the $B$–$X$ cross term in the field strengths and yields canonically normalized kinetic terms. In fact, the choice \eqref{eq:kin_mix_transform} is exact: substituting $\hat B_\mu = B_\mu - \frac{\epsilon}{\sqrt{1-\epsilon^2}}X_\mu$ and $\hat X_\mu = \frac{1}{\sqrt{1-\epsilon^2}}X_\mu$ into the kinetic Lagrangian, the kinetic mixing matrix is diagonalized to the identity.  
In other words
the mixed field--strength term is removed by the following non--orthogonal transformation in the basis $(B,W^3,X)$:
\begin{equation}
\begin{pmatrix}
\hat{B}_\mu \\[4pt] \hat{W^3}_\mu \\[4pt] \hat{X}_\mu
\end{pmatrix}
\;=\;
G(\epsilon)\,
\begin{pmatrix}
B_\mu \\[4pt] W^3_\mu \\[4pt] X_\mu
\end{pmatrix},
\qquad
G(\epsilon) \;=\;
\begin{pmatrix}
1 & 0 & -\dfrac{\epsilon}{\sqrt{1-\epsilon^2}}\\[6pt]
0 & 1 & 0\\[6pt]
0 & 0 & \dfrac{1}{\sqrt{1-\epsilon^2}}
\end{pmatrix}.
\end{equation}

We will henceforth work in the basis of fields $(B_\mu,\;W^3_\mu,\;X_\mu)$, which has standard kinetic terms. In this basis, the interaction terms also transform accordingly. Notably, the new gauge field $X_\mu$ now couples not only to the $L_i-L_j$ current but also acquires a tiny coupling to the hypercharge current (proportional to $\epsilon$). One way to see this is to substitute the inverse of \eqref{eq:kin_mix_transform} into the current interactions in \eqref{eq:Lag_start}. We find: 
\[ 
-\,g'\,J_Y^\mu \hat B_\mu \;-\; g_X\,J_X^\mu \hat X_\mu \;\;=\;\; -\,g'\,J_Y^\mu B_\mu \;-\; g_X\,J_X^\mu X_\mu \;+\; \epsilon\,g'\,J_Y^\mu X_\mu \;+\; \mathcal{O}(\epsilon^2)~,
\] 
where we have dropped an overall normalization factor $ \sqrt{1-\epsilon^2}\approx 1$ on the $X$ coupling for simplicity. The last term shows that after removing kinetic mixing, $X_\mu$ couples to the hypercharge current with strength $\epsilon g'$. Equivalently, $X_\mu$ picks up a small coupling to the electromagnetic current (since the hypercharge current contains the electromagnetic current component, as we will see later). 

Next we consider the mass terms for the neutral gauge bosons. The SM $SU(2)_L\times U(1)_Y$ is spontaneously broken to $U(1)_{\text{EM}}$ by the Higgs doublet vacuum expectation value (VEV) $v=\langle H^0\rangle \approx 246~\mathrm{GeV}$. The neutral Higgs current is $(D_\mu H)^\dagger (D^\mu H)$ which gives mass to a particular linear combination of $B_\mu$ and $W^3_\mu$. In the $(B,\,W^3)$ subspace (ignoring $X$ for the moment), the mass-squared matrix arising from the Higgs VEV is 
\begin{equation}
\label{eq:SM_mass_matrix}
M_{\text{SM}}^2 \;=\; \frac{v^2}{4}\begin{pmatrix} g'^2 & -\,g\,g' \\[6pt] -\,g\,g' & g^2 \end{pmatrix}~. 
\end{equation}
The matrix \eqref{eq:SM_mass_matrix} has a zero eigenvalue corresponding to the photon and a nonzero eigenvalue giving the $Z$-boson mass. Diagonalizing this $2\times2$ submatrix is accomplished by the usual electroweak (Weinberg) mixing angle $\theta_W$, defined by $\tan\theta_W = \frac{g'}{\,g\,}$. The orthonormal eigenstates are:
\begin{equation}
\label{eq:weinberg_rotation}
A_{0\mu} = \cos\theta_W\, B_\mu + \sin\theta_W W^3_\mu , 
\qquad\qquad 
Z_{0\mu} = - \sin\theta_W B_\mu + \cos\theta_W W^3_\mu ,
\end{equation}
where $A_{0\mu}$ is the neutral massless state (the photon before mixing with $X$) and $Z_{0\mu}$ is the massive state (the $Z$ boson state prior to $X$-mixing). In this convention, the tree-level $Z$ mass is 
\[ 
m_{Z_0}^2 \;=\; \frac{v^2}{4}\,\big(g^2 + g'^2\big)~, 
\] 
and the photon is massless ($m_{A_0}=0$). We note for later convenience that the electromagnetic coupling is $e = g\sin\theta_W = g'\cos\theta_W$, and we define the useful combination $g_Z \equiv \sqrt{g^2 + g'^2} = \frac{g}{\cos\theta_W} = \frac{g'}{\sin\theta_W}$, which is the coupling constant of the $Z$ boson to the $Z$-current.

Now we incorporate the new gauge boson $X_\mu$ into the mass matrix. We assume that the $U(1)_{L_i-L_j}$ symmetry is spontaneously broken (for example, by a new Higgs field carrying that charge or via a Stueckelberg mechanism), giving the $X_\mu$ boson a mass $m_X$. Importantly, the SM Higgs is neutral under $L_i-L_j$, so it does not directly induce any mixing mass term between $X_\mu$ and the electroweak bosons. In the original gauge basis $(\hat B,\hat W^3,\hat X)$, this means that prior to kinetic mixing there were no off-diagonal mass terms involving $\hat X$. However, after transforming to the canonical $(B,W^3,X)$ basis via \eqref{eq:kin_mix_transform}, small off-diagonal mass terms \emph{do} appear at $\mathcal{O}(\epsilon)$. Intuitively, this is because the combination of fields that couples to the Higgs (namely $B_\mu$) now contains a small admixture of the $X_\mu$ field. Concretely, substituting $B_\mu = \hat B_\mu + \epsilon \hat X_\mu$ into the Higgs mass terms, one finds new cross-terms between $X$ and $B,W^3$. At leading order in $\epsilon$, the full $3\times3$ neutral boson mass-squared matrix in the basis $(B_\mu,\;W^3_\mu,\;X_\mu)$ can be written as 
\begin{equation}\label{eq:mass_matrix_full}
\mathcal{M}^2 \;\approx\; \frac{v^2}{4}\begin{pmatrix}
g'^2 & -\,g\,g' & -\,g'^2\,\epsilon \\[6pt]
-\,g\,g' & g^2 & g\,g'\,\epsilon \\[6pt]
-\,g'^2\,\epsilon & g\,g'\,\epsilon & \displaystyle \frac{4\,m_\chi^2}{v^2}
\end{pmatrix}~,
\end{equation}
up to corrections of $\mathcal{O}(\epsilon^2)$ (which are negligible for small mixing). In writing the $(3,3)$ element above, we have factored out $v^2/4$ for convenience; equivalently, one can read $4m_\chi^2/v^2$ as $(m_\chi^2 / m_{Z_0}^2)\,(g^2+g'^2)$ since $m_{Z_0}^2=(g^2+g'^2)v^2/4$. The off-diagonal entries proportional to $\epsilon$ in \eqref{eq:mass_matrix_full} are the induced mass mixings: note in particular the $B$–$X$ term (1,3) and $W^3$–$X$ term (2,3).

We can now generalize the diagonalization in the basis of $B, W^3$ reported in Eq.~\ref{eq:weinberg_rotation} adding also the boson $X$:
\begin{equation}
R_W(\theta_W)=
\begin{pmatrix}
 c_W & s_W & 0\\
 -s_W &  c_W & 0\\
 0   &   0  & 1
\end{pmatrix},
\qquad
\begin{pmatrix}
A_{0\mu}\\[2pt]Z_{0\mu}\\[2pt]X_\mu
\end{pmatrix}
=
R_W\,\begin{pmatrix}B_\mu\\ W^3_\mu\\ X_\mu\end{pmatrix},
\label{eq:Rw}
\end{equation}
Now that we have diagonalized the $(B,W^3)$ sub-block via the rotation \eqref{eq:weinberg_rotation}, we now express the remaining mass mixing between the $Z_0$ and $X$ states. In the basis $(A_{0\mu},\,Z_{0\mu},\,X_\mu)$, the photon $A_{0}$ is still an eigenstate with zero mass, and the $2\times2$ mass matrix for the $(Z_0,\,X)$ sector is 
\begin{equation}\label{eq:Z0X_matrix}
\begin{pmatrix} 
m_{Z_0}^2 & \Delta^2 \\[6pt] \Delta^2 & m_\chi^2 
\end{pmatrix}~, 
\end{equation}
where $m_{Z_0}^2 = \frac{v^2}{4}(g^2+g'^2)$ as above, and $\Delta^2$ is the off-diagonal mixing term. To find $\Delta^2$, we project out the $Z_0$ component from the $B$–$X$ and $W^3$–$X$ entries of \eqref{eq:mass_matrix_full}. Using $Z_{0\mu} = \cos\theta_W\,W^3_\mu - \sin\theta_W\,B_\mu$ (see \eqref{eq:weinberg_rotation}), one finds 
\[ 
\Delta^2 \;=\; \epsilon\,\sin\theta_W\,m_{Z_0}^2~.
\] 
In other words, the off-diagonal mass term is proportional to the kinetic mixing parameter $\epsilon$ and to $\sin\theta_W$ (which reflects the fact that hypercharge mixing induces $Z$ mixing only in proportion to the $B_\mu$ content of the $Z_0$ field). The physical mass eigenstates $Z$ and $X$ are obtained by a further rotation of angle $\xi$ (sometimes called the ``dark mixing angle'') in the $Z_0$–$X$ plane. Specifically, we define 
\begin{equation}\label{eq:dark_rotation}
\begin{pmatrix} Z_\mu \\[3pt] X_\mu \end{pmatrix} \;=\; \begin{pmatrix} \cos\xi & \sin\xi \\[3pt] -\,\sin\xi & \cos\xi \end{pmatrix} \begin{pmatrix} Z_{0\mu} \\[3pt] X_\mu \end{pmatrix}~,
\end{equation} 
chosen such that all mixing is removed from the mass matrix. The angle $\xi$ is given by 
\begin{equation}\label{eq:tan2xi}
\tan(2\xi) \;=\; \frac{2\,\Delta^2}{\,m_\chi^2 - m_{Z_0}^2\,} \;=\; \frac{2\,\epsilon\,\sin\theta_W\,m_{Z_0}^2}{\,m_\chi^2 - m_{Z_0}^2\,} = \frac{2\,\epsilon\,\sin\theta_W}{\,1 - \delta\,}. 
\end{equation}
where we have defined for convenience $\delta \equiv m_\chi^2/m_{Z_0}^2$. 
Assuming $\epsilon/(1-\delta)\ll 1$ we can approximate:
\begin{equation}
\label{eq:approxmix}
\sin\xi \approx \frac{\epsilon \sin\theta_W}{(1-\delta)}\, , \, \cos \xi \approx 1 .
\end{equation}
This is typically true for most of the cases since $\epsilon$ is constrained to be smaller than $10^{-3}$ from collider experiments \cite{Cline:2024qzv}.
The only region in the parameter space where the above approximations are not correct is when $m_\chi \sim m_{Z_0}$ for which even a small $\epsilon$ can induce a significant mixing angle $\xi$. In what follows we will keep $\xi$ as a free parameter determined by \eqref{eq:tan2xi}, without assuming a hierarchy between $m_\chi$ and $m_{Z_0}$.

We can thus define in the basis $(W^3,B,X)$ another matrix that now diagonalize the $B-X$ part:
\begin{equation}
R_X(\xi)=
\begin{pmatrix}
1 & 0 & 0\\
0 &  \cos{\xi} &  \sin{\xi}\\
0 & -\sin{\xi} &  \cos{\xi}
\end{pmatrix},\qquad
\tan2\xi=\frac{2\Delta^2}{m_X^2-m_Z^2}
=\frac{2\epsilon \sin \theta_W}{1-\delta},
\label{eq:Rxi}
\end{equation}

After performing the rotation \eqref{eq:dark_rotation}, we arrive at the mass eigenstate basis $\{A_\mu,\;Z_\mu,\;A'_\mu\}$. The physical massless photon field is $A_\mu = A_{0\mu}$ (since the photon had no mixing with $X_\mu$ and remains an eigenstate). The massive $Z_\mu$ and $A'_\mu$ fields are given in terms of the original electroweak and $X$ fields by combining \eqref{eq:weinberg_rotation} and \eqref{eq:dark_rotation}: 
\begin{equation}\label{eq:physical_basis}
\begin{pmatrix} A_\mu \\[3pt] Z_\mu \\[3pt] A'_\mu \end{pmatrix} \;=\; 
\begin{pmatrix}
\cos\theta_W & \sin\theta_W & 0 \\[3pt]
-\sin\theta_W\cos\xi & \,\cos\theta_W\cos\xi & \sin\xi \\[3pt]
\sin\theta_W\sin\xi & -\,\cos\theta_W\sin\xi & \,\cos\xi 
\end{pmatrix}
\begin{pmatrix} B_\mu \\[3pt] W^3_\mu \\[3pt] X_\mu \end{pmatrix}~,
\end{equation}
where we have written the full rotation matrix $R_1(\xi)R_2(\theta_W)$ acting on the original basis $(W^3, B, X)$ for completeness. The state $Z_\mu$ is identified as the SM $Z$ boson (slightly modified by mixing), and $A'_\mu$ is the new heavy neutral gauge boson. The eigenvalues corresponding to $Z$ and $X$ can be obtained by diagonalizing \eqref{eq:Z0X_matrix}. In closed form, the $Z$ and $A'$ masses are 
\begin{equation}\label{eq:Z_Zp_masses}
m_{Z,A'}^2 \;=\; \frac{1}{2}\Big[\,m_{Z_0}^2 + m_\chi^2 \;\pm\; \sqrt{(m_X^2 - m_{Z_0}^2)^2 + 4\,\Delta^4}\,\Big]~,
\end{equation}
with $\Delta^2$ given above. Expanding to first order in the small parameters $\epsilon,\xi$, one finds $m_Z \approx m_{Z_0}\,\big(1 + \frac{\epsilon^2\sin^2\theta_W}{2(1-\delta)}\big)$ and $m_{A'} \approx m_X\,\big(1 + \frac{\epsilon^2\sin^2\theta_W}{2(\delta-1)}\big)$, so the shifts in the masses are of order $\epsilon^2$ and typically negligible for our purposes. In summary, $A_\mu$ remains massless, $Z_\mu$ has mass $m_Z \approx m_{Z_0}$, and $A'_\mu$ has mass $m_{A'} \approx m_X$. We now turn to the interaction eigenstates of these bosons and derive the interaction Lagrangian in the mass basis.

The charged currents (involving $W^\pm$) are not affected by the new $U(1)$ and follow the usual form; here we focus on the neutral-current interactions. It is convenient to introduce the electromagnetic current $J_{\rm EM}^\mu$ and the SM $Z$-boson current $J_Z^\mu$:
\begin{equation}\label{eq:currents_EM_Z}
J_{\rm EM}^\mu \;=\; \sum_f Q_f\,\bar f \gamma^\mu f~, \qquad\qquad 
J_Z^\mu \;=\; \sum_f \Big(T_3^f - \sin^2\theta_W\,Q_f\Big)\,\bar f \gamma^\mu f~,
\end{equation}
where $Q_f$ is the electric charge of fermion $f$ (in units of $e$), and $T_3^f$ its $SU(2)_L$ third-component. 
The fermionic currents $J_{\rm EM}$ and $J_Z$ are defined as:
\begin{equation}
J_{\rm EM}^\mu=J_3^\mu+J_Y^\mu,\qquad
J_Z^\mu=J_3^\mu-s_W^2J_{\rm EM}^\mu
\label{eq:currents}
\end{equation}

By construction, the photon couples to $J_{\rm EM}^\mu$ and the SM $Z_0$ boson couples to $J_Z^\mu$ (often called the neutral weak current). Meanwhile, the new $U(1)$ gauge field couples to the $L_i-L_j$ current $j_X^\mu$. We can now write the neutral-current part of the interaction Lagrangian after transforming into the mass eigenstate basis. Starting from \eqref{eq:Lag_start} and using the rotations \eqref{eq:weinberg_rotation} and \eqref{eq:dark_rotation}, one finds at all orders in $\epsilon$ and $\xi$:
\begin{eqnarray}
\mathcal{L}_{\text{int}} &=&
-e\,A_\mu J_{\text{EM}}^{\mu}
-
Z_\mu\!\left[
     g_Z\!\left(\cos\xi + \frac{s_W\,\epsilon\,\sin\xi}{\sqrt{1-\epsilon^2}}\right) J_Z^{\mu}
   \;-\; e\,\frac{\cos \theta_W\,\epsilon\,\sin\xi}{\sqrt{1-\epsilon^2}}\,J_{\text{EM}}^{\mu}
   \;+\; g_X\,\frac{\sin\xi}{\sqrt{1-\epsilon^2}}\,J_X^{\mu}
\right]
+ \\
&-& A'_\mu\!\left[
     g_X\,\frac{\cos \xi}{\sqrt{1-\epsilon^2}}\,J_X^{\mu}
   \;-\; e\,\frac{\cos \theta_W\,\epsilon\,\cos\xi}{\sqrt{1-\epsilon^2}}\,J_{\text{EM}}^{\mu}
   \;+\; g_Z\!\left[\frac{\sin \theta_W\,\epsilon\,\cos\xi}{\sqrt{1-\epsilon^2}} -
         \sin\xi
       \right] J_Z^{\mu}
\right]. \nonumber
\end{eqnarray}


Instead, at leading order in $\epsilon$ and $\xi$ and taking $\cos \xi \approx 1$ the previous equation simplifies as: 
\begin{align}\label{eq:L_int_massbasis}
\mathcal{L}_{\text{int}}  = & - e A_\mu J_{\rm EM}^\mu \notag \\[5pt]
&{}- Z_\mu \Big[g_Z \;J_Z^\mu + g_X \sin\xi  J_X^\mu \Big] \notag \\[3pt]
&{}- A'_\mu \Big[g_X\,  J_X^\mu \;-\; e \epsilon \cos\theta_W  J_{\rm EM}^\mu  + g_Z( \epsilon \sin \theta_W  -  \sin\xi )J_Z^\mu \Big]~,
\end{align}
up to negligible higher-order terms $\mathcal{O}(\epsilon^2,\xi^2)$. 

The structure of \eqref{eq:L_int_massbasis} is very informative. The first line is the photon coupling: $A_\mu$ couples exclusively to the electromagnetic current with strength $e$, as expected. The second line gives the couplings of the physical $Z$ boson. The term proportional to $g_Z$ is the SM $Z$ coupling to the neutral current $J_Z^\mu$ (up to the small normalization factor $\cos\xi\approx1$). However, we see one additional piece: the $g_X\,\sin\xi\,J_X^\mu$ term shows that the $Z$ boson also gains a small coupling to the new $L_i-L_j$ current, suppressed by the mixing angle $\xi$. Likewise, the third line of \eqref{eq:L_int_massbasis} gives the couplings of the new $A'$ boson. It couples predominantly to the $L_i-L_j$ current $j_X^\mu$ with strength $g_X\cos\xi\approx g_X$. But it also has a millicharged coupling to the electromagnetic current, $-\epsilon e \cos\theta_W J_{\rm EM}^\mu$ (which allows $A'$ to interact feebly with electrically charged particles even if they carry no $L_i-L_j$ charge). And finally, the $X$ couples to the SM neutral current $J_Z^\mu$ with a suppressed strength $g_Z(\epsilon \sin \theta_W-\sin\xi)$. In particular, for $\delta = m_\chi/m_Z \ll 1$, $\epsilon \sin \theta_W-\sin\xi \approx 0$ (see Eq.~\ref{eq:approxmix}) and the mixing between $A'$ and the $J_Z$ is negligible. 
In summary, due to kinetic and mass mixing, the physical $X$ boson can interact with ordinary SM fermions (which may have $q_{X,f}=0$) through a combination of small $\epsilon$- and $\xi$-suppressed couplings, even though those fermions had no direct coupling to the gauge eigenstate $\hat X_\mu$.

\end{document}